 \documentclass[preprint,review,12pt]{elsarticle}

\usepackage{amssymb}

\usepackage{amsmath}
\usepackage{multicol}
\usepackage{multirow}
\usepackage{tabularx}
\usepackage{float}  
\usepackage{subfloat}  
\usepackage{graphicx, subfig}
\usepackage{caption}
\usepackage{dcolumn}
\usepackage{bm}
\usepackage{rotating}
\usepackage{booktabs}

\newcommand{\beqa}{\begin{eqnarray}}
\newcommand{\eeqa}{\end{eqnarray}}
\newcommand{\beq}{\begin{equation}}
\newcommand{\eeq}{\end{equation}}
\newcommand{\be}{\begin{equation*}}
\newcommand{\ee}{\end{equation*}}
\newcommand{\bi}{\begin{itemize}}
\newcommand{\ei}{\end{itemize}}
\newcommand{\ben}{\begin{enumerate}}
\newcommand{\een}{\end{enumerate}}

\newcommand{\dif}{\mathrm{d}}

\begin{document}

\begin{frontmatter}



\title{A large eddy simulation (LES) study of inertial deposition  of
  particles onto in-line tube-banks}


\author{C.~Jin,
  ~I.~Potts,~D. C.~Swailes,
  M. W.~Reeks\footnote{Email:mike.reeks@ncl.ac.uk}}
\address{School of Mechanical \& Systems Engineering, Newcastle University, Stephenson Building,
Claremont Road, Newcastle upon Tyne, NE1 7RU, UK}

\date{\today}

\begin{abstract}
  We study deposition and impact of heavy particles onto an in-line
  tube-banks within a turbulent cross flow through Lagrangian particle tracking coupled with an LES modelling
  framework. The flow Reynolds number based on the cylinder
  diameter $D$ and flow velocity between the gap of two
  vertically adjacent cylinders is $33960$. We examine the
  flow structures across the tube bank and report surface
  pressure characteristics on cylinders. Taking into account
  particle-wall impact and bounce, we study dispersion and
  deposition of three sets of particles ($St = 0.35, 0.086,
  0.0075$) based on $10^7$ particles tracked through the
  turbulent flow resolved by LES. The deposition efficiency
  for the three sets of particles are reported across the
  tube-banks. Positions of particle deposited onto
  tube-banks shows that significantly more of
  smaller particles deposit onto the back-side of the
  back-banks. This suggests that the smaller
  particles are easier to be entrained into the wake and
  impact onto the back-side of cylinders.

\end{abstract}


\end{frontmatter}



\section{Introduction}
Investigation of the deposition and impact of aerosol
particles on heat exchangers is of significant importance to
the design and operation of heat exchanger tube banks used
in a wide range of industrial applications, i.e., civil
advanced gas-cooled reactor (CAGR) boilers, oil-fired steam
boilers of thermal power stations and process plants.  In
many safety cases involving dropped fuel in CAGRs a
significant proportion of the activity will be associated
with small aerosol particles. The main mechanisms by which
aerosol particulates deposit and impact on wall surfaces
include gradient/diffusion or free-flight theory , inertia
deposition, interception, turbulent eddy-diffusion, Brownian
diffusion and thermophoresis, gravitational setting, etc.  The
mechanisms that are responsible for the deposition of heavy
particles in fully developed turbulent boundary layers are
gradient/diffusion or free flight theory
~(\cite{Friedlander1957}) and by turbophoresis
~(\cite{reeks1983transport}) with turbulent eddy-diffusion
~(\cite{Kallio1989}). However, under the conditions of high
volume flow rate low pressure drop filtration, inertial
impact becomes the dominant mechanism governing deposition
among all the competing ones that contribute to the
deposition of aerosol particulates on cylinder surfaces
(see~ \cite{helgesen1991particle,helgesen1994particle}).

There has been extensive research regarding the inertia
deposition of heavy particles or droplets from flowing gas
streams by impact on a single cylinder surfaces through
theory and experiments.  For example,
\citeauthor{brun1955impingement} \cite{brun1955impingement} reported three impingement
characteristics of water droplets on a cylinder surface,
which are total rate of water droplet impingement, extend of
droplet impingement zone and local distribution of impinging
water on cylinder surface. The results on the collection
efficiency of a cylinder in \citeauthor{brun1955impingement} \cite{brun1955impingement} were
presented as a function of combining the Stokes and Reynolds
number of droplets considered. This treatment was extended
to use a generalized Stokes number to determine the
collection efficiency of a cylinder for non-Stokesian
particles by \citeauthor{israel1982use}\cite{israel1982use}. This generalized Stokes
number is normally referred to as effective Stokes number
and defined as 
\beq St_{\mathrm{eff}} = \psi(Re_p) St, \eeq
where $\psi(Re_p)$ is the non-Stokes drag correction factor
and given by 
\beq \psi(Re_p) = \frac{24}{Re_p}\int_0^{Re_p}
\frac{1}{C_D(Re^{\prime})\,Re^{\prime}}\,\dif Re^{\prime},
\label{eq:st_psi}
\eeq
and 
\beq
St = \frac{\rho_p d_p^2 U_o}{18 u D}.
\eeq 
More recently, the collection efficiency was examined
through directly solving the incompressible Navier-Stokes
equations coupled with the Lagrangian point particle
tracking approach in a relatively low Reynolds number cross
flow across a cylinder by \citeauthor{haugen2010particle}\cite{haugen2010particle}.

There are a number of numerical studies on the deposition
and impact of heavy particles on tube-banks surfaces, and
they focused on the two-dimensional simulations.
\citeauthor{tabakoff1994numerical}\cite{tabakoff1994numerical} carried out a two-dimensional
numerical simulation for a dilute particle laden laminar
flow over in-line tube-banks in order to study particle
impact and erosion of cylinders. Rebound phenomena of
particles from cylinder surfaces were take into account as
well in the above work. \citeauthor{bouris2001numerical}\cite{bouris2001numerical} performed
a two-dimensional large eddy simulation to evaluate
alternate tube configurations for particle deposition rate
reduction on heat exchanger tube bundles, in which an energy
balance model was implemented to consider the adhesion or
rebound of particles upon hitting upon the surface a
tube. \citeauthor{tian2007numerical}\cite{tian2007numerical} made use of the
two-dimensional RANS (Reynolds-averaged Navier-Stokes)
modelling framework and Lagrangian particle tracking to study the characteristics of particle-wall
collisions. An algebraic particle-wall collision and
stochastic wall roughness model was also implemented by
\citeauthor{tian2007numerical} \cite{tian2007numerical}.

\begin{figure}[htp]
  \centering
  \includegraphics[trim=3.0in 0.0in 3.0in
  1.0in,clip=true,scale=1.0, angle = 0, width = 1.0\textwidth]{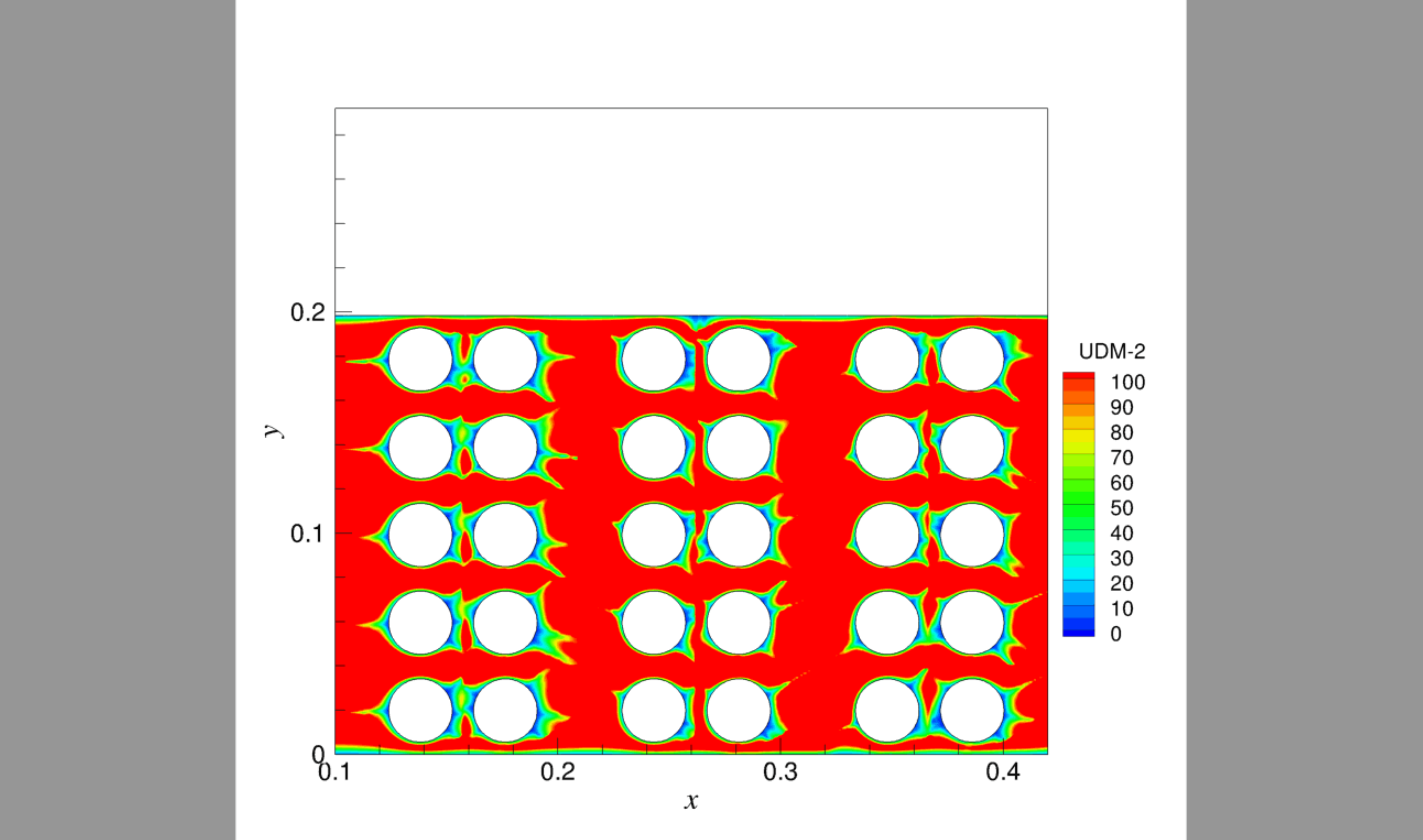} 
  \caption[Contour of the $y^{+}$ value of each cell
  associated with its correspondingly nearest wall-adjacent
  cell face in a turbulent flow across in-line
  tube-banks. UDM-2 stands for $y^{+}$.]  {Contour of the
    $y^{+}$ value of each cell associated with its
    correspondingly nearest wall-adjacent cell face in a
    turbulent flow across in-line tube-banks. UDM-2 stands
    for $y^{+}$.}
  \label{fig:urans_yplus}
\end{figure}

Engineering predictions of the deposition of heavy particles
on bluff bodies depends primarily on the RANS. The
methodology of three-dimensional RANS modelling frameworks
coupled with a separate boundary layer model, which supplies
fluctuating fluid velocity fluctuations seen by heavy
particles, has been extended to study the prediction of
deposition rates of heavy particles (e.g. \cite{Dehbi2008a,jin2015simple})
in complex geometries. \citeauthor{dehbi2011cfd}\cite{dehbi2011cfd} further employed
above mentioned method to study particulate flows around
linear arrays of spheres and got good predicted deposition
rates when compared against experimental
measurements. However, for a turbulent flow across bluff
bodies, e.g. spheres or cylinders, the salient feature of
such a flow is that it has strongly unsteady,
three-dimensional vortex shedding
(\cite{williamson1996vortex}). This requires solving the
Navier-Stokes equations with the time-dependent term,
i.e. unsteady RANS (URAS) or LES, in order to resolve the
unsteady phenomena of vortex shedding as accurately as
possible. In this study, first a URAS simulation was carried
out for a turbulent flow across in-line tube-banks. The
approach presented in \citeauthor{Dehbi2008a}\cite{Dehbi2008a} was used to
determine the $y^{+}$ value of each cell associated with its
correspondingly nearest wall-adjacent cell face. However, as
shown in figure~\ref{fig:urans_yplus} for the contour of
$y^{+}$ values of each cell associated with its
correspondingly nearest wall-adjacent cell face, the
boundary layer around every cylinder based on a threshold
$y^{+}$ value $100$ possesses a irregular shape. This
irregular boundary layer shape as a result of the
unsteadiness of vortex shedding may make the methodology of
RANS modelling framework combining with a supplying boundary
model problematic.

LES has been convincingly demonstrated to be superior to
unsteady RANS (URANS) in accurately predicting the flow and
vortex dynamics of a turbulent cross-flow in a staggered (\cite{benhamadouche2003coarse}) and in-line
tube bundle (\cite{jin2016study}). This is because
LES is capable of providing the detailed large scales of
flow structures, resolving a significant part of the vortex
shedding physics and hence reducing the importance of
modelling. The success of LES technique for single-phase
turbulent flows across complex geometries has been explored
to extend the technique to two-phase flows over complex
geometries. \citeauthor{apte2003large}\cite{apte2003large} performed an LES study of
particle-laden swirling flow in a coaxial-jet
combustor. They demonstrated that results obtained from LES
are significantly more accurate than the results by RANS
applied for the same problem. \citeauthor{Riber2009}\cite{Riber2009} conducted a
comparison study of numerical strategies for LES of
particulate two-phase recirculating flows and observed that
the dispersed phase is predicted more accurately by the
Lagrangian point particle approach than the Eulerian
approach. Therefore, the Lagrangian point particle approach
coupled with the LES technique is employed to in this study.

The principal objective of this work is to investigate
inertial deposition and impact of heavy particles onto
in-line tube-banks in a turbulent cross flow. The numerical
technique used for the underlying flow field is large eddy
simulation (LES), whilst the Lagrangian point particle tracking
approach is employed to obtain particles trajectories.

\section{Overview of numerical simulations}

\subsection{Formulation of a dynamic Smagorinsky model}
The governing equations for LES are obtained by spatially
filtering the Navier-Stokes equations. In this process, the
eddies that are smaller than the filter size used in the
simulations are filtered out. Hence, the resulting filtered
equations govern the dynamics of large eddies in turbulent
flows. A spatially filtered variable that is denoted by an
overbar is defined using a convolution product
(see~\cite{leonard1974energy})
\beq
\overline{\phi}(\mathbf{x}, t) =
\underset{\mathcal{D}}\int %
\phi(\mathbf{y},t)G(\mathbf{x}, \mathbf{y}) d\mathbf{y}
\eeq
where $\mathcal{D}$ denotes the computational domain, and
$G$ the filter function that determines the
scale of the resolved eddies.

In the current study, the finite-volume discretization
employed itself provides the filtering operation as
\beq
\overline{\phi}(\mathbf{x}, t) = \frac{1}{\mathit{V}}\underset{\mathcal{D}}\int %
\phi(\mathbf{y}, t)d\mathbf{y}, \quad \mathbf{y} \in
\mathcal{V}
\label{filtering}
\eeq
where $\mathit{V}$ denotes the volume of a computational
cell. Hence, the implied filter function, $G(\mathbf{x},
\mathbf{y})$ in eq.\eqref{filtering}, is a top-hat filter
given by
\beq
G(\mathbf{x}, \mathbf{y}) = \begin{cases}
  1/\mathit{V} & \quad \text{for } |\mathbf{x} - \mathbf{y}| \in
  \mathcal{V} \\
  0 & \quad \text{otherwise}
  \end{cases}
\eeq
Filtering the continuity and Navier-Stokes  equations, the
governing equations for resolved scales in LES are obtained
\beq
\frac{\partial{\overline{u}}_i}{\partial{x}_i} = 0
\eeq
\beq
\frac{\partial{\overline{u}_i}}{\partial{t}} +
\frac{\partial{\overline{u}_i\overline{u}_j}}{\partial{x}_j}
= -\frac{1}{\rho}\frac{\partial{\overline{p}}}{\partial{x}_i} +
\frac{\partial}{\partial{x}_j}
\left ( \nu\frac{\partial{\overline{u}_i}}{\partial{x}_j}
\right ) - \frac{\partial{\tau_{ij}}}{\partial{x}_j}
\eeq
where $\tau_{ij}$ denotes the subgrid scale (SGS herefrom) stress tensor
defined by
\beq
\tau_{ij} = \overline{u_i u_j} - \overline{u}_i
\overline{u}_j
\label{sgs0}
\eeq
The filtered equations are unclosed since the SGS stress
tensor $\tau_{ij}$ is unknown.  The SGS stress tensor can
be  modelled based on
an isotropic eddy-viscosity model as:
\beq
\tau_{ij} -
\frac{1}{3}\tau_{kk} \delta_{ij} =
-2\nu_{t}\overline{S}_{ij}
\eeq
where  $\nu_t$ denotes the SGS eddy viscosity, and
$\overline{S}_{ij}$ is the resolved rate of strain tensor given
by
\beq
\overline{S}_{ij} = \frac{1}{2}\left (
  \frac{\partial{\overline{u}_i}}{\partial{x}_j} +
  \frac{\partial{\overline{u}_j}}{\partial{x}_i} \right )
\eeq
where $\nu_t$ is computed in terms of the \citeauthor{smagorinsky1963general}
\cite{smagorinsky1963general} type eddy-viscosity model
using
\beq
\nu_t = C_{\nu} \overline{\Delta}^2 |\overline{S}|
\eeq
where $C_\nu$ denotes the Smagorinsky coefficient, $|\overline{S}|$
 the modulus of rate of strain tensor for the resolved scales,
\beq
|\overline{S}| = \sqrt{2\overline{S}_{ij}\overline{S}_{ij}}
\eeq
and $\overline{\Delta}$ denotes the grid filter length obtained from
\beq
\overline{\Delta} = \mathit{V}^{1/3}
\eeq
Consequently, the SGS stress tensor is computated as
following
\beq
\tau_{ij} - \frac{1}{3}\delta_{ij} \tau_{ij} =
-2C_{\nu}\overline{\Delta}^2|\overline{S}|\overline{S}_{ij}
\label{smag1}
\eeq
This model claims to be simple and efficient.  It needs
merely a constant \textit{in priori} value for
$C_{\nu}$. Nevertheless, work from
\cite{lilly1966application,deardorff1970numerical,piomelli1988model}
has shown different values of $C_{\nu}$ for distinct
flows. Hence, the major drawback of the model used in LES is
that there is an inherent inability to represent a wide
range of turbulent flows with a single value of the model
coefficient $C_{\nu}$. Given that the turbulent flow over
tube-banks in the present study is fully three-dimensional,
the standard Smagorinsky SGS model is not used here to
compute the coefficient $C_{\nu}$.

\citeauthor{germano1991dynamic}\cite{germano1991dynamic} proposed a new procedure to
dynamically compute the model coefficient $C_{\nu}$ based
on the information obtained from the resolved large scales
of motion. The new procedure employes another coarser filter
$\widetilde{\Delta}$ (test filter)
whose width is greater than that of the default grid
filter. Applying the test filter to the filtered
Navier-Stokes equations, one obtains the following
equations
\beq
\frac{\partial{\widetilde{\overline{u}}_i}}{\partial{t}} +
\frac{\partial{\widetilde{\overline{u}}_i}\widetilde{\overline{u}}_j}{\partial{x}_j}
= -\frac{1}{\rho}\frac{\partial{\widetilde{\overline{p}}}}{\partial{x}_i} +
\frac{\partial}{\partial{x}_j}
\left ( \nu\frac{\partial{\widetilde{\overline{u}}_i}}{\partial{x}_j}
\right ) - \frac{\partial{\mathit{T}_{ij}}}{\partial{x}_j}
\eeq
where the tilde denotes the test-filtered
quantities. $\mathit{T}_{ij}$ represents the subgrid scale
stress tensor from the resolved large scales of motion and is given
by
\beq
\mathit{T}_{ij} = \widetilde {\overline{u_i u_j}} - \widetilde{\overline{u}}_i\widetilde{\overline{u}}_j
\label{sgs1}
\eeq
The quantities given in \eqref{sgs0} and \eqref{sgs1} are
related by the Germano identity:
\beq
\mathcal{L}_{ij} = \mathit{T_{ij}} - \widetilde{\tau}_{ij}
\label{germano}
\eeq
which represents the resolved turbulent stress tensor from
the SGS tensor between the test and grid
filters,$\mathit{T}_{ij}$ and $\tau_{ij}$. Applying the same
Smagorinsky model to $\mathit{T}_{ij}$ and $\tau_{ij}$, the
anisitropic parts of $\mathcal{L}_{ij}$ can be written
as
 \beq
 \mathcal{L}_{ij} - \frac{1}{3}\mathcal{L}_{kk}\delta_{ij} = -2CM_{ij}
 \eeq
 where
 \beq
 M_{ij} =
 \widetilde{\Delta}^2 |\widetilde{\overline{S}}|\widetilde{\overline{S}}_{ij}
 -
 \overline{\Delta}^2\widetilde{|\overline{S}|\overline{S}}_{ij}
 \label{germano1}
 \eeq
 One hence obtains the value of $C$ from \eqref{germano1}
 that is solved on the test filter level and then apply it to Eq.
 \eqref{smag1}.  The model value of  $C$ is
 obtained via a least squares approach proposed by
 \citeauthor{lilly1992proposed}\citep{lilly1992proposed}, since Eq. \eqref{germano1} is an
 overdetermined system of equations for the unknown variable $C$.  
\citeauthor{lilly1992proposed}\cite{lilly1992proposed} defined a criterion for minimizing
the square of the error as
\beq
 E = (L_{ij} - \frac{\delta_{ij}}{3}L_{kk} - 2CM_{ij})^2
\eeq
In order to obtain a local value, varying in time and space
in a fairly wide range, for the model constant $C$, one
takes $\frac{\partial{E}}{\partial{C}}$ and sets it zero to get
\beq
C = \frac{1}{2}\frac{L_{ij}M_{ij}}{M_{ij}M_{ij}}
\eeq
A negative C represents the transfer of flow energy from the
subgrid-scale eddies to the resolved eddies, which is known
as \textit{back-scatter} and
regarded as a desirable attribute of
the dynamic model.

\subsection{The Werner and Wengle wall layer model}
The Large Eddy Simulation (LES) of turbulent flow over tube-banks
is hampered by expensive computational cost incurred when the
dynamic and thin near-wall layer is fully resolved. To
obviate the computational cost associated with calculating
the wall shear stress from the laminar stress-strain relationship that
requires the first cell to be put within the range of $y^{+}
\approx 1$, \citeauthor{werner1993large}\cite{werner1993large} proposed a simple power-law to
replace the law of the wall, in which the velocity
profile on a solid wall is given as following,
\beq
u^{+} = \begin{cases}
  y^{+} & \quad \text{for } y^{+} \le 11.81\\
  A(y^+)^B & \quad \text{for } y^{+} > 11.81
\end{cases}
\label{ww-model}
\eeq
where $A = 8.3$ and $B=1/7$. An analytical integration of Eq.
\eqref{ww-model} results in the following relations for the
wall shear stress
\beq
|\tau_w| =
\begin{cases}
  \frac{2\mu|u_p|}{\Delta y} & \quad \text{for } y^{+} \le 11.81\\
  \rho \left [ \frac{1- B}{2}A^{\frac{1+B}{1-B}}\left
      (\frac{\mu}{\rho \Delta y}\right )^{1+B} 
     + \frac{1 + B}{A} \left( \frac{\mu}{\rho \Delta
      y}^B |u_p|\right )\right ]^{\frac{2}{1+B}}  & \quad \text{for } y^{+} > 11.81
\end{cases}
\label{ww-model}
\eeq
where $u_p$ is velocity component parallel to the wall and given by:
\beq
|u_p| = \frac{\mu}{2\rho\Delta y}A^{\frac{2}{1-B}}
\eeq

\subsection{Flow configuration of in-line tube banks}
\begin{figure}[htp]
  \centering
  \includegraphics[trim=0.2in 0.0in 0.2in
  0.0in,clip=true,scale=1.0, angle = 0, width = 1.0\textwidth]{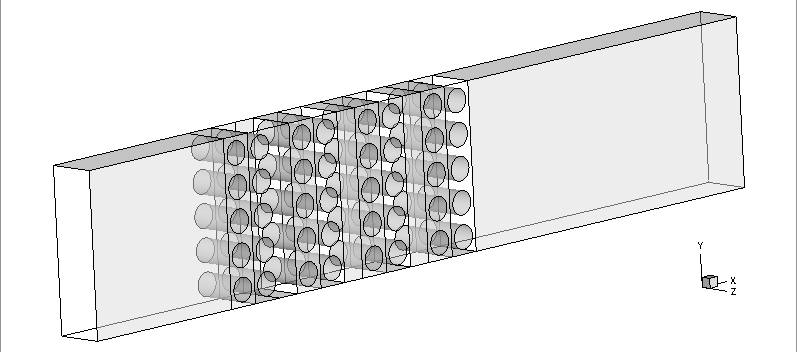} 
  \caption[Configuration of four by five pairs of in-line tube
banks.]
{Configuration of four by five pairs of in-line tube
banks.} 
   \label{fig:be_tubebank}
\end{figure}

Figure~\ref{fig:be_tubebank} shows the flow configuration
with the corresponding coordinate system, which is based on
the experiments involving particle deposition on heat
exchanger tube-banks by \citeauthor{hall1994}\cite{hall1994}. Flow direction is from left
to right and normal to the cylinder axis. The computational
domain is of size $L_x \times L_y \times L_z = 36.16D \times
6.94D \times 2D$, where $D$ is the cylinder diameter. It
consists of four by five pairs of in-line tube
banks. Every pair has the transverse pitch that is of the
ratio of pitch-to-diameter ($P_T/D $) $S_T$ equalling to
1.388 and the longitudinal pitch that is of the ratio of
pitch-to-diameter ($P_L/D$) $S_{L0}$ equalling to 1.331,
respectively. The longitudinal pitch between the two
adjacent cylinders from two adjacent tube-banks is of the
ratio of pitch-to-diameter ($P_L/D$) $S_{L1}$ equalling to
2.331.  The Reynolds number $Re_o$based on the free stream
velocity $U_o$ and the cylinder diameter $D$ equals to
$9500$, and consequently in terms of the continuity equation
$Re_g$ that is based on the gap velocity in the $x-$
direction between two cylinders equals to $33960$.

\begin{figure}[htp]
  \centering
  \includegraphics[trim=0.2in 0.0in 0.2in
  0.0in,clip=true,scale=1.0, angle = 0, width = 1.0\textwidth]{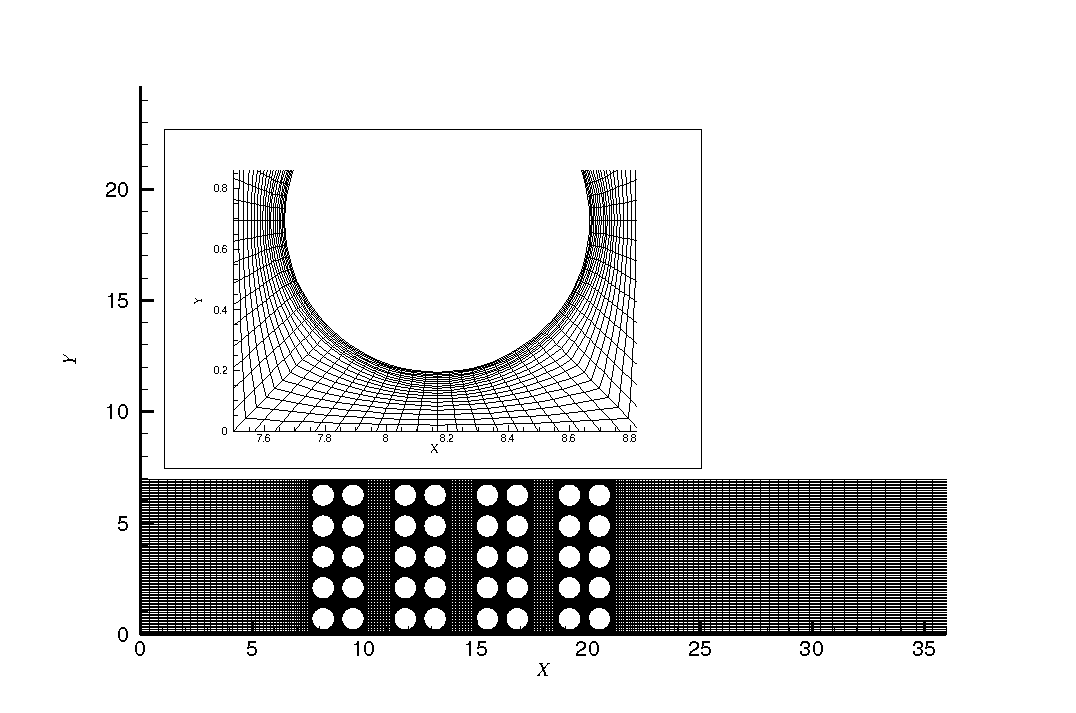} 
  \caption[A side view of computational mesh used for the
  LES with a close look-up around a cylinder.]
{A side view of computational mesh used for the
  LES with a close look-up around a cylinder.}
   \label{fig:mesh_be_tubebank}
\end{figure}

Figure~\ref{fig:mesh_be_tubebank} shows a side view of the
computational grid with a close look-up around a cylinder.
The total number of grid elements used for the present
simulation is around $3.4$ million. The mesh has an embedded
region of fine mesh designed for each cylinder in order to
enhance the mesh resolution near the cylinder without
incurring too large an increase in the total number of mesh
elements.  The first cell adjacent to the cylinder is within
the range $\Delta y^{+} < 11.8$ in wall units\footnote{The
  superscript $+$ denotes a non-dimensional quantity scaled
  using the wall variables, e.g. $y^{+} = yu_{\tau}/\nu$,
  where $\nu$ is the kinematic viscosity and $u_\tau =
  \sqrt{\tau_w/\rho}$ is the wall friction velocity based on
  the wall shear stress ${\tau}_w$, and which is a velocity
  scale representative of velocities close to a solid
  boundary.}, which satisfies the requirements of the
Wener-Wengle wall-layer model used for wall-modelling
LES. Prior to the present simulation, with the standard
Smagorinsky subgrid scale model, a simulation based coarse
grid resolution were carried out to determine the
resolution.

With fully developed turbulent flows, periodic boundary
conditions are justified to use along the normal ($y$) and
spanwise ($z$) direction. For the inlet boundary condition,
a simple uniform velocity profile is assumed and the
turbulent intensity set to zero. Hence, the turbulence
fluctuations at the inlet was not accounted for temporally
and spatially. Nevertheless, a length $7.5D$ before the
first column bank is used to allow the development of
turbulence. At the exit boundary, the solution variables
from the adjacent interior cells are extrapolated to satisfy
the mass conservation.

The simulation is advanced with non-dimensional time step
$\Delta t U_o/D \approx 1.4 \times 10^{-3}$ that yields
maximum Courant-Friedrichs-Lewy (CFL) number 0.7. For the
carrier phase, the first-order statistics are collected by
integrating the governing equations over an interval of
$25D/U_o$, and all the statistics are averaged over the $40$
sampling points across in the spanwise direction.

\subsection{Calculation of particle trajectories}
A parallel Lagrangian particle tracking module was developed
to calculate
trajectories of heavy particles in flow fields resolved by LES. The particle
localization algorithm on unstructured grids proposed by
\citeauthor{haselbacher2007efficient}\cite{haselbacher2007efficient}
was used to locate the cell which contains the current
particle position. In this study, the focus is on the
deposition of non-inter-collision, rigid, spherical and
heavy particles on in-line tube-banks by impact; the
concentration of particles is dilute enough to make one-way
coupling assumption. The momentum balance equation of
particles discussed by
\citeauthor{maxey1983equation}\cite{maxey1983equation} is simplified in this
work with taking into account the drag force merely. We thus
can write the particle equation of motion involving the
non-linear form of the drag law with the point particle
approximation \beq \frac{\dif \mathbf{u}_p}{\dif t} =
\frac{1}{\tau_p}C_D\frac{Re_p}{24}(\mathbf{u} -
\mathbf{u}_p),
\label{eq:ch7bem}
\eeq 
where $\mathbf{u_p}$ is the particle
velocity and $\mathbf{u}$ the instantaneous fluid velocity
at the particle location, $\tau_p$ is the particle response
time. An empirical relation for $C_D$ from \citeauthor{Morsi1972}\cite{Morsi1972},
which is applicable to a wide range of particle Reynolds
number with sufficiently high accuracy, is employed.
\beq 
C_D = c_1 + \frac{c_2}{Re_p} +
\frac{c_3}{Re_p^2},
\label{eq:Cdexpression}
\eeq
in which $c_1, c_2, c_3$ are
constants. The above empirical expression exhibits the correct
asymptotic behavior at low as well as high values of
$Re_p$.  

The position $\mathbf{x}_p$ of particles is obtained from the kinematic
relationship
\beq
\frac{\dif \mathbf{x}_p}{\dif t} = \mathbf{u}_p
\label{eq:ch7kinematicrelation}
\eeq
The boundary condition for the above equation is that the
particle is captured by the wall when its center away the
nearest wall surface is less than its radius. This is not
properly treated in the default discrete phase model (DPM)
provided by ANSYS FLUENT. Furthermore, this error has a
significant effect upon predictions concerning the
deposition of heavy particles under investigation.

From a statistically stationary LES flow field, equation:
(\ref{eq:ch7kinematicrelation}) is started to be integrated
in time using the second-order Adams-Bashforth scheme to get
particle trajectories, whilst equation: (\ref{eq:ch7bem}) is
integrated with the second-order accurate Gear2 (backward
differentiation formulae) scheme that is applicable to stiff
systems. Since it is only by chance that a particle
coincides with the cell centroid, at which the fluid
solution is stored as part of the computation of the
underlying flow field with the unstructured-grid based
collocated cell centroid storage finite volume method, a quadratic scheme based on
least-squares velocity gradient reconstruction is used to
interpolate the fluid velocity to the particle location
without consideration of the effect of sub-grid scale. For
the second-order accurate carrier phase solver,
the quadratic scheme is found to be sufficient to accurately
resolve the particle motions.

Results on for three sets of particles ($St = 0.35, 0.086, 0.0075$) are obtained by following the trajectories of $10^7$
particles which are continuously released into the
computational domain. This large number of particles
trajectories are crucial in order to present statistically
significant results on the particles phase which interacts
with the vortex dynamics in spatial and temporal
domain. However, the unsteady deposition issue of particles
on the tube-banks has not been taken into account, which
results in a significant simplification on the interaction
of incoming particles with deposition particles.

\subsection{A particle-wall collision model}
\begin{figure}[htp]
  \centering
  \includegraphics[trim=0.0in 0.0in 0.0in
  0.0in,clip=true,scale=1.0, angle = 0, width = 0.95\textwidth]{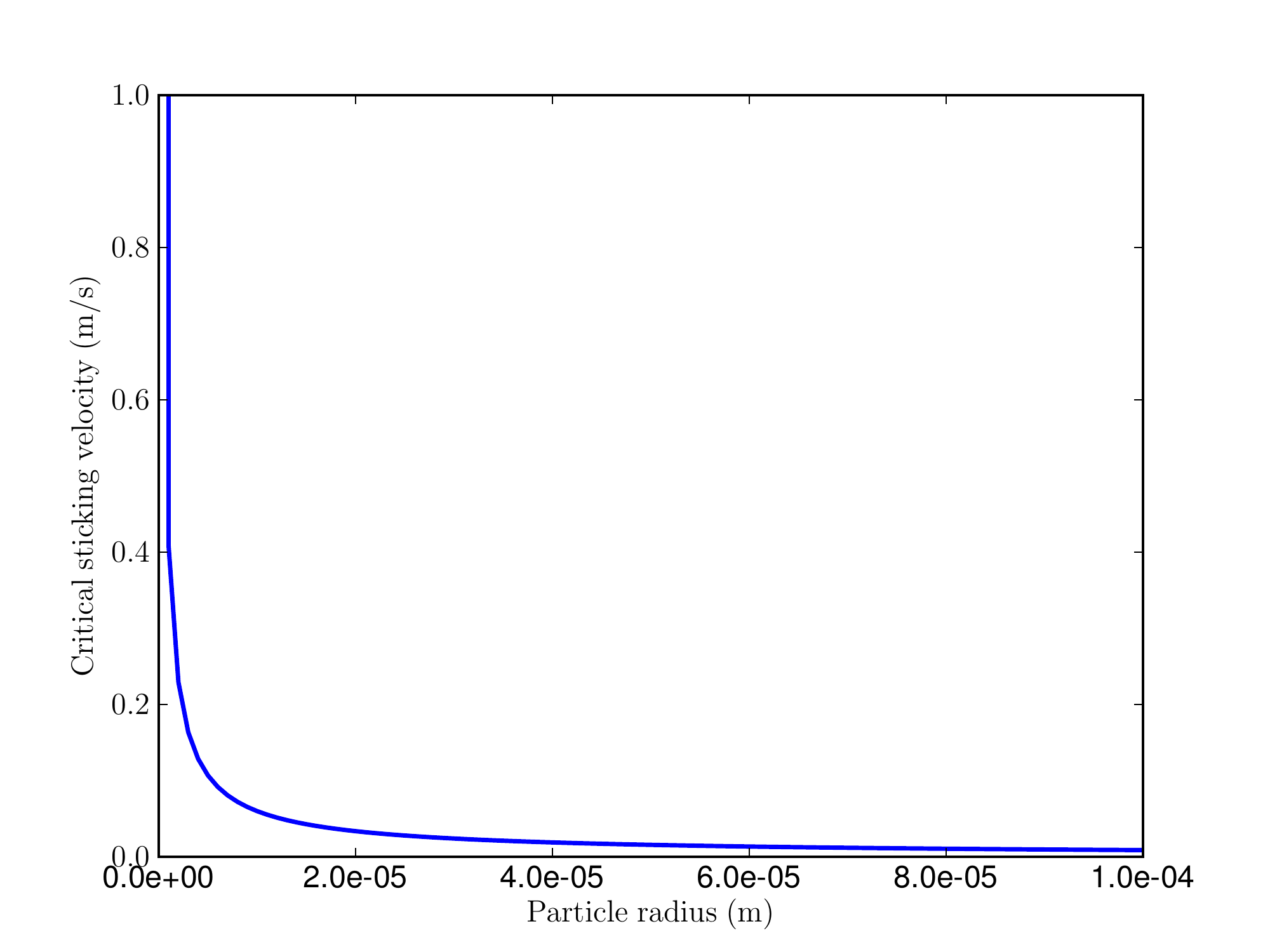} 
  \caption[Critical sticking velocity as a function of particle radius.]
{Critical sticking velocity as a function of particle radius.} 
   \label{fig:be_critical_sticking}
\end{figure}

\begin{figure}[htp]
  \centering
  \includegraphics[trim=0.0in 0.0in 0.0in
  0.0in,clip=true,scale=1.0, angle = 0, width = 0.95\textwidth]{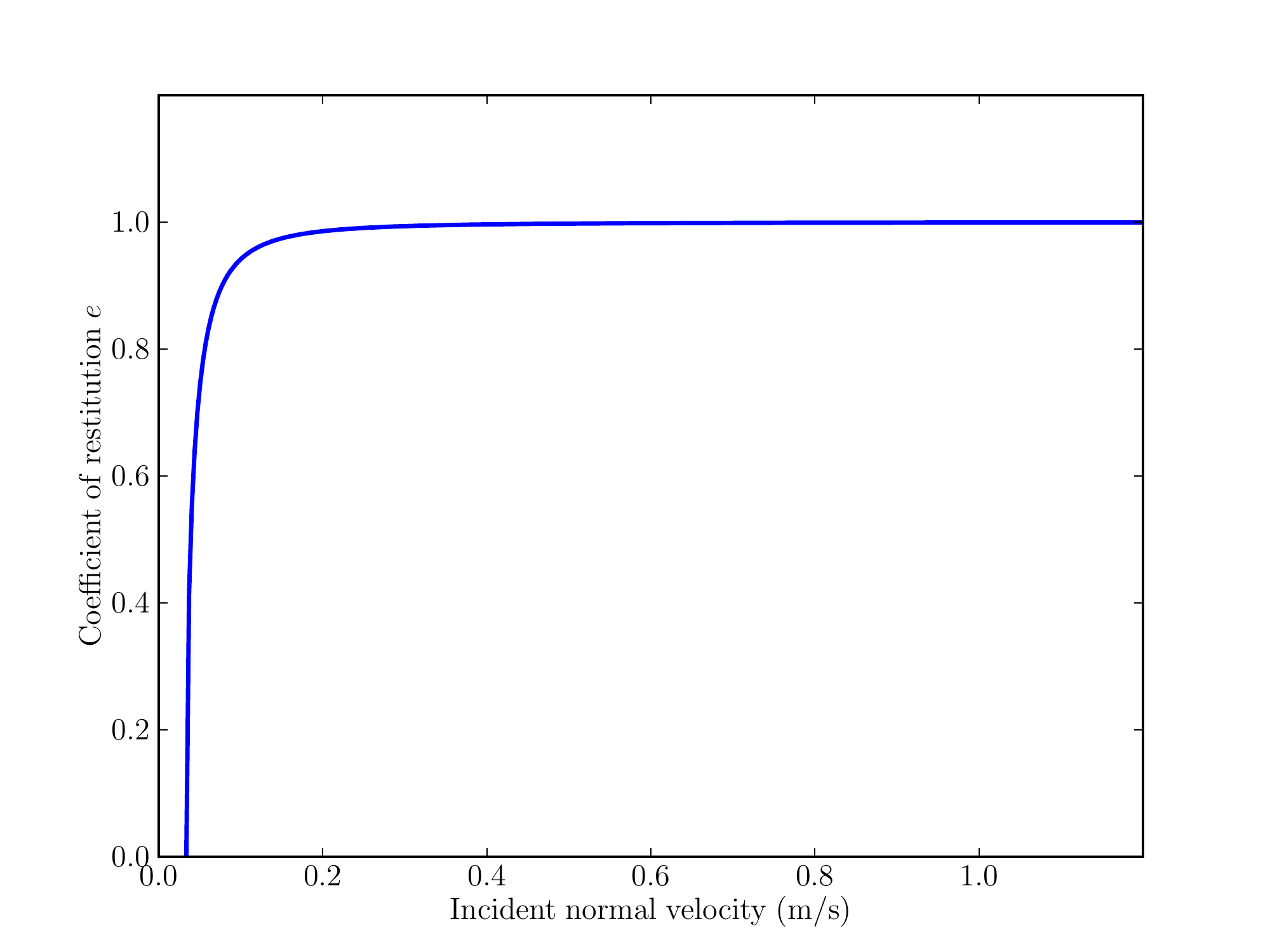} 
  \caption[Coefficient of restitution $e$ as a function of particle
  incident normal velocity.]
{Coefficient of restitution $e$ as a function of particle
  incident normal velocity.} 
   \label{fig:be_CoR}
\end{figure}

Particle-wall collisions play an important role in
particle-laden two-phase flows because it effects the
deposition, accumulation on wall surfaces. In this work, the
aim is not to seek a new particle-wall-collision model,
instead a well-known dry particle-wall-collision model from
\citeauthor{thornton1998theoretical}\cite{thornton1998theoretical} was implemented to account
for the energy loss resulting from the
particle-wall-collision. The energy loss resulting from
impact upon a wall is normally characterized by the
coefficient of restitution (CoR) $e$ that is defined by
\beq 
e = \frac{v_r^{n}}{v_i^{n}} 
\eeq 
where $v_{r}^{n}$ is the rebound normal velocity and $v_{i}^{n}$ the incident normal
velocity. Then, the loss of kinetic energy $\Delta E$ of a
particle with mass $m_p$ is given by
\beq
\Delta E = \frac{1}{2}m_p \left ( {v_{i}^{n}}^2 - {v_{r}^{n}}^2
\right) = \frac{1}{2}m_p {v_i^n}^2 \,\left(1 - e^2 \right). 
\eeq
In the case of elastic impact, $e = 1$ due to no
energy loss occurred. When $e = 0$, the maximum incident
normal velocity is normally referred to as the critical
sticking velocity $v_{s}$. Then from
\beq
\Delta E = \frac{1}{2} m_p v_s^2,
\eeq 
$e$ is given by
\beq
e = \left[1 - \left(\frac{v_s}{v_i^{n}} \right)^2 \right].
\eeq
If $v_{i}^{n}$ is higher than $v_{s}$ then
$e > 0$ and the particle can bounce off the wall upon impact; if not, the particle
sticks the wall and $e = 0$.  The critical sticking velocity $v_s$ is normally determined
by the properties of the particle and
wall. Figure~\ref{fig:be_critical_sticking} shows the
variation of critical sticking velocity for a wide range of
particle radii. It can be observed that the smaller the
particle radius is, the larger the critical sticking
velocity is. Therefore, it is easier for larger particles
get bounce upon impact. Figure~\ref{fig:be_CoR} illustrates
how the coefficient of restitution with the particle
incident normal velocity. When the particle incident normal
velocity is approaching $0.2~\mathrm{m/s}$ the coefficient
of restitution is close to $0.985$.

In this study, the critical sticking velocities for three
sets of particles considered are calculated and input as
parameters before starting particle tracking. Therefore,
this model can be used to determine whether a particle
sticks to or rebound from a wall upon impact with the wall.

\section{Results and discussions}

\subsection{Results on the carrier phase}
\begin{figure}[htp]
  \centering
  \includegraphics[trim=0.0in 1.0in 2.0in
  1.8in,clip=true,scale=1.0, angle = 0, width = 0.95\textwidth]{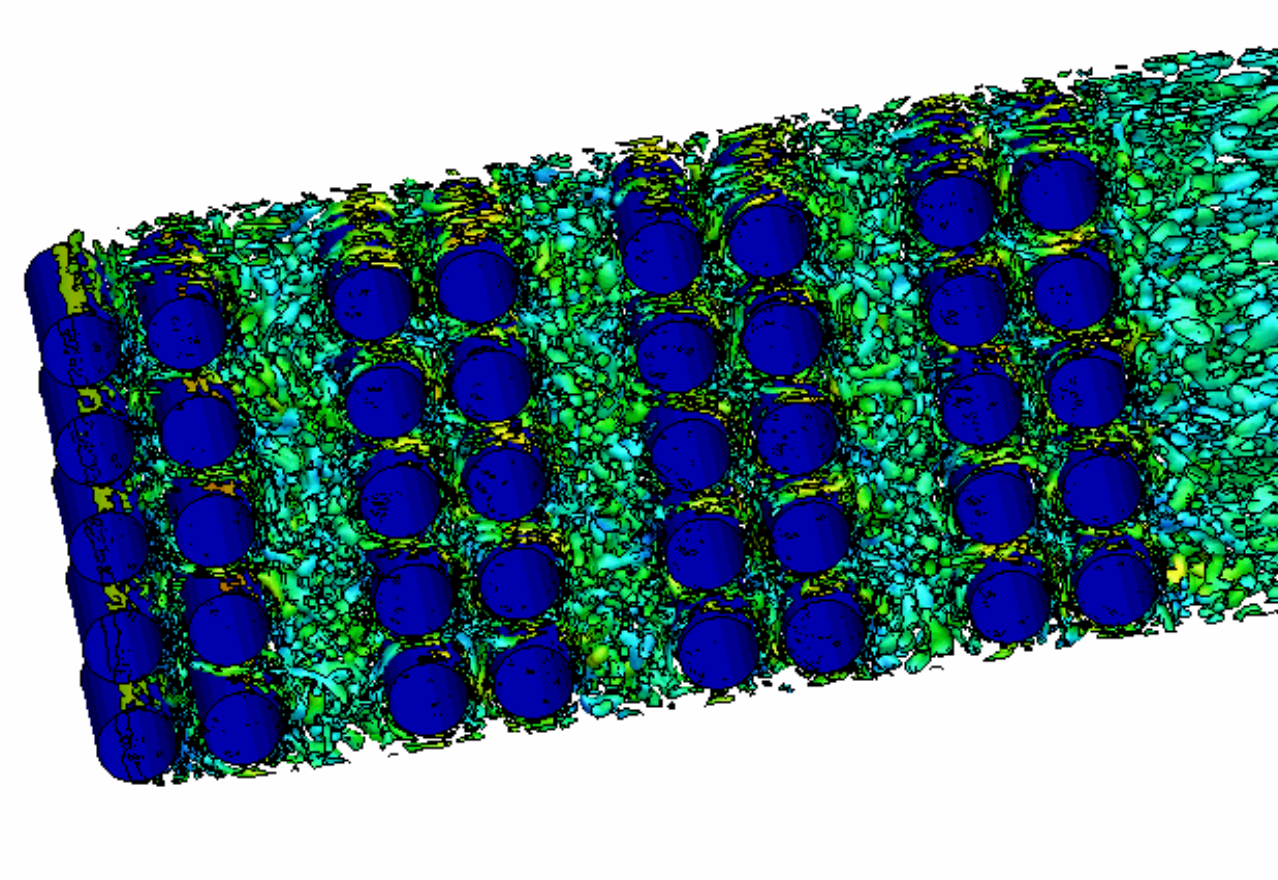} 
  \caption[Instantaneous velocity magnitude based on the
  normalized Q criterion equalling to $0.08$.]
{Instantaneous velocity magnitude based on the
  normalized Q criterion equalling to $0.08$.} 
   \label{fig:be_tb_Q}
\end{figure}

Figure~\ref{fig:be_tb_Q} shows the contour of the
instantaneous velocity magnitude based on the normalized Q
criterion equalling to 0.08. The top
cylinder in the first column develops a laminar boundary
layer and has some kind of laminar vortex shedding. However, this
was not observed from the downstream cylinders. Large
coherent structures are visible in the gaps
between tube-banks, but they are not as well organized and
periodic as in typical Karman vortex streets for a single
cylinder at the similar Reynolds number. Large coherent
structures between two adjacent column cylinders in the same
pair are not as obvious as those in the gaps. This may
result from the relatively small gap in every pair destroys
the development of the wake. Finally, the flow is evolving
into a turbulent flow like a grid turbulence from the final
pair of tube-banks.

Following \citeauthor{shim1988fluctuating}\cite{shim1988fluctuating}, the coefficient for the mean pressure
distribution on the cylinder surface is define as
\beq
\overline{C}_p = \frac{\left< p \right>_{T} -
  p_{ref}}{q_{ref}},
\label{eq:cp_ch7}
\eeq
where $\left< p \right>_{T}$ denotes an ensemble average across
the spanwise direction for all the sampling points on the
cylinder surface over the
sampling time interval $T$, and
\beq 
 q_{ref} = \frac{1}{2}\rho u_g^2. 
\eeq
In order to make $\overline{C}_p$ equal to unit at the front
stagnation point for every cylinder, the corresponding
static pressure $p_{ref}$ is calculated according to
equation \ref{eq:cp_ch7}, $\overline{C}_p$ is hence
determined around the cylinder surface. 

\begin{figure}[htp]
\centering
\subfloat[][]{
\includegraphics[width=0.95\textwidth]{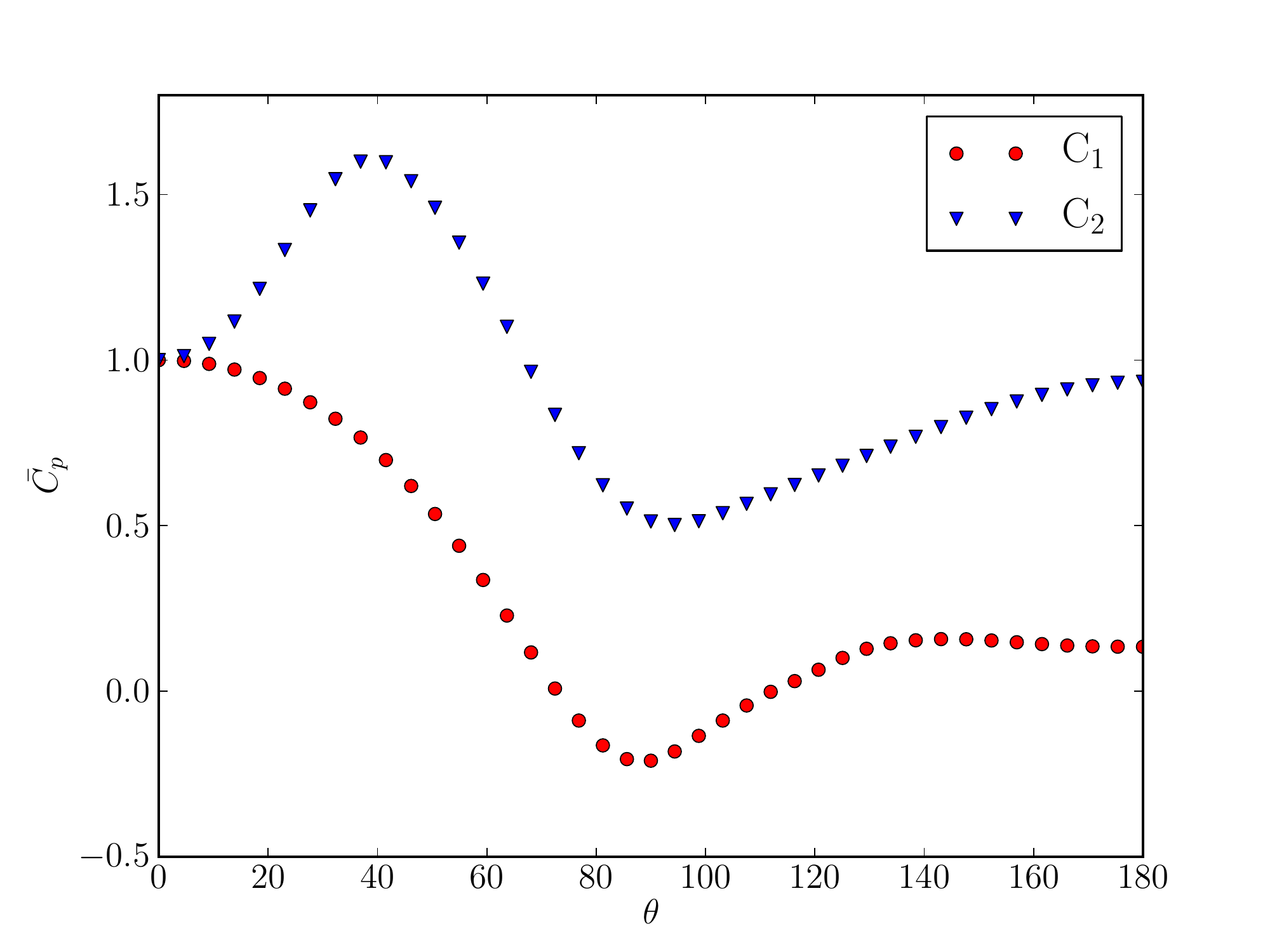}
\label{fig:subfig1_Cp_be}}\\
\subfloat[][]{
\includegraphics[width=0.95\textwidth]{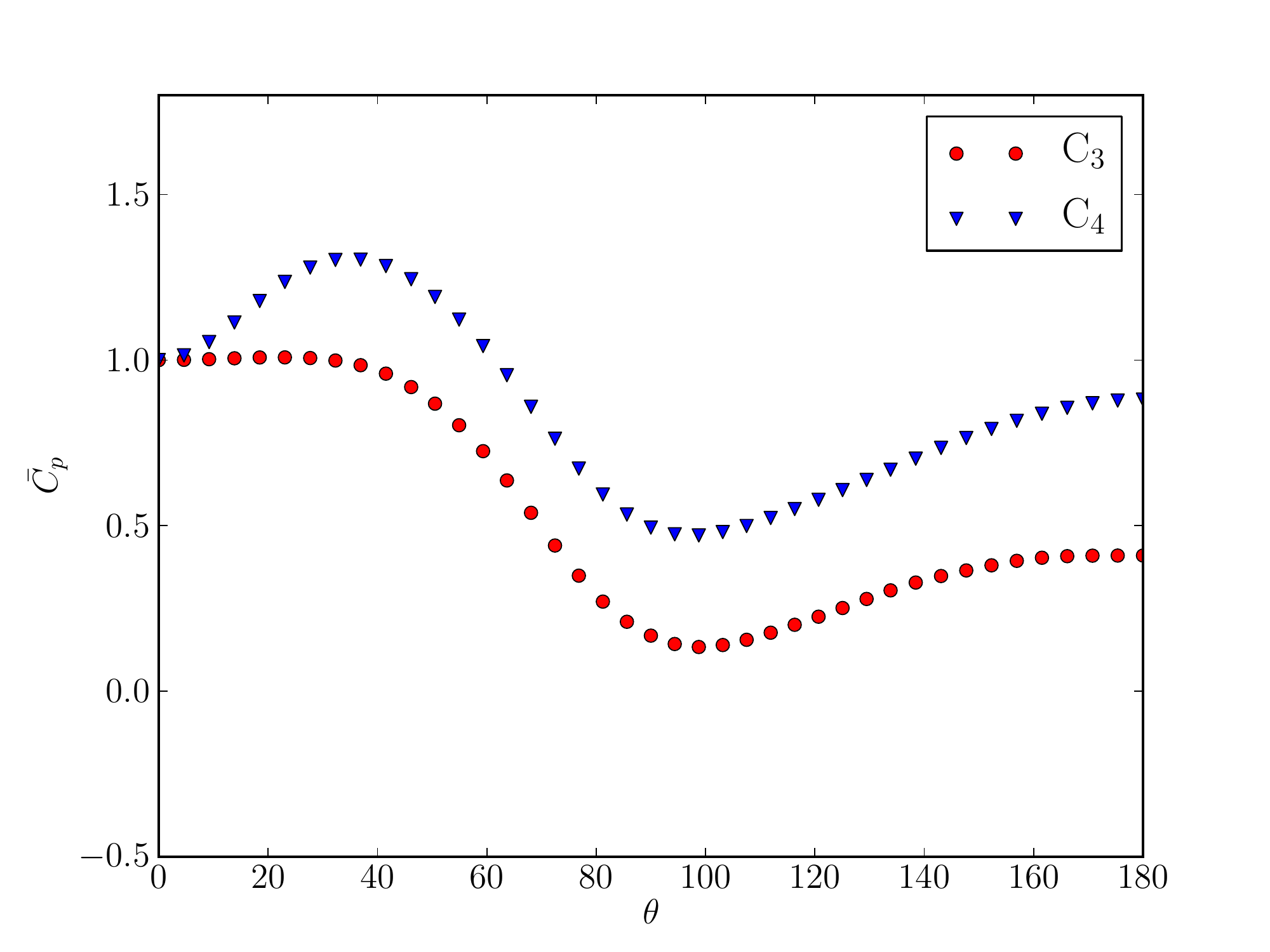}
\label{fig:subfig2_Cp_be}}
\caption[Mean pressure distribution on the middle cylinder
  surface. Definition of $\overline{C}_p$
  based on \cite{shim1988fluctuating}. (a) the first pair
  tube banks, (b) the second pair
  tube banks.]
{Mean pressure distribution on the middle cylinder
  surface. Definition of $\overline{C}_p$
  based on \cite{shim1988fluctuating}. (a) the first pair
  tube banks, (b) the second pair
  tube banks.}
\label{fig:globfig1_Cp_be}
\end{figure}

Comparisons of $\overline{C}_p$ on the middle cylinders
surface from the first and second pair of tube-banks are
shown in figure \ref{fig:globfig1_Cp_be}. It can be observed
that $\overline{C}_p$ is of the standard shape, which
$\overline{C}_p$ normally develops on a single circular
cylinder.  However, $\overline{C}_p$ on the second cylinder
is of a $\mathbf{S}$ shape, indicating there is a region in
the front side of the cylinder that has higher pressure than
the front stagnation point. This higher pressure
distribution results from the impingement of the wake from
the preceding cylinder on the front side. The phenomena was
also observed in the experimental measurements for a
tube-banks with a close longitudinal pitch from \citeauthor{shim1988fluctuating}
\cite{shim1988fluctuating}.  From~\ref{fig:subfig2_Cp_be},
It can be observed that $\overline{C}_p$ on $C_3$ is also of
the standard shape for a single cylinder, but it has a
relatively high base pressure. $\overline{C}_p$ on the
following cylinder $C_3$ has the similar shape like to the
one on $C_4$, implying that the shedding vortex from $C_3$
impacts on the front side of $C_4$.

\begin{figure}[htp]
\centering
\subfloat[][]{
\includegraphics[width=0.95\textwidth]{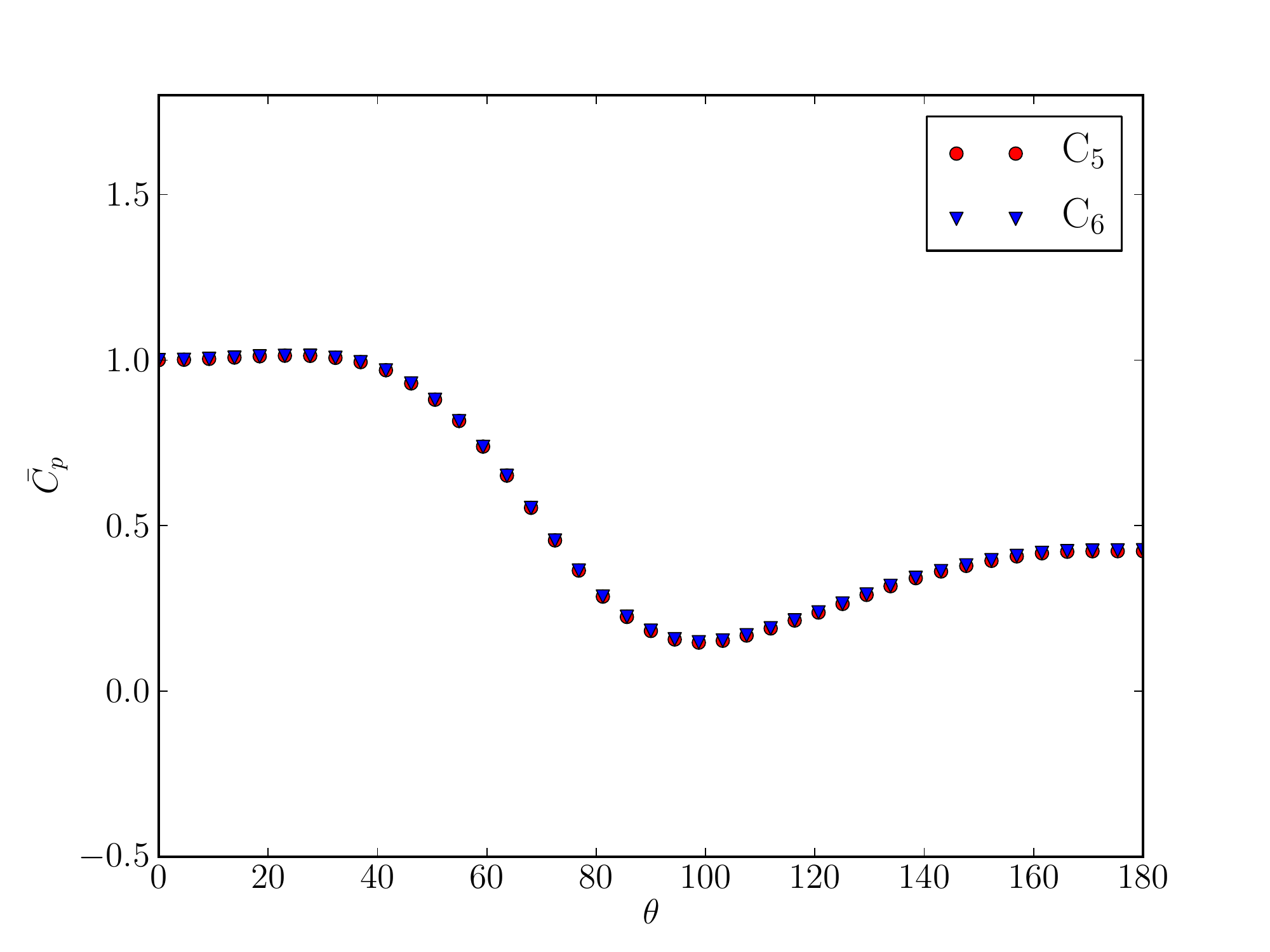}
\label{fig:subfig3_Cp_be}}\\
\subfloat[][]{
\includegraphics[width=0.95\textwidth]{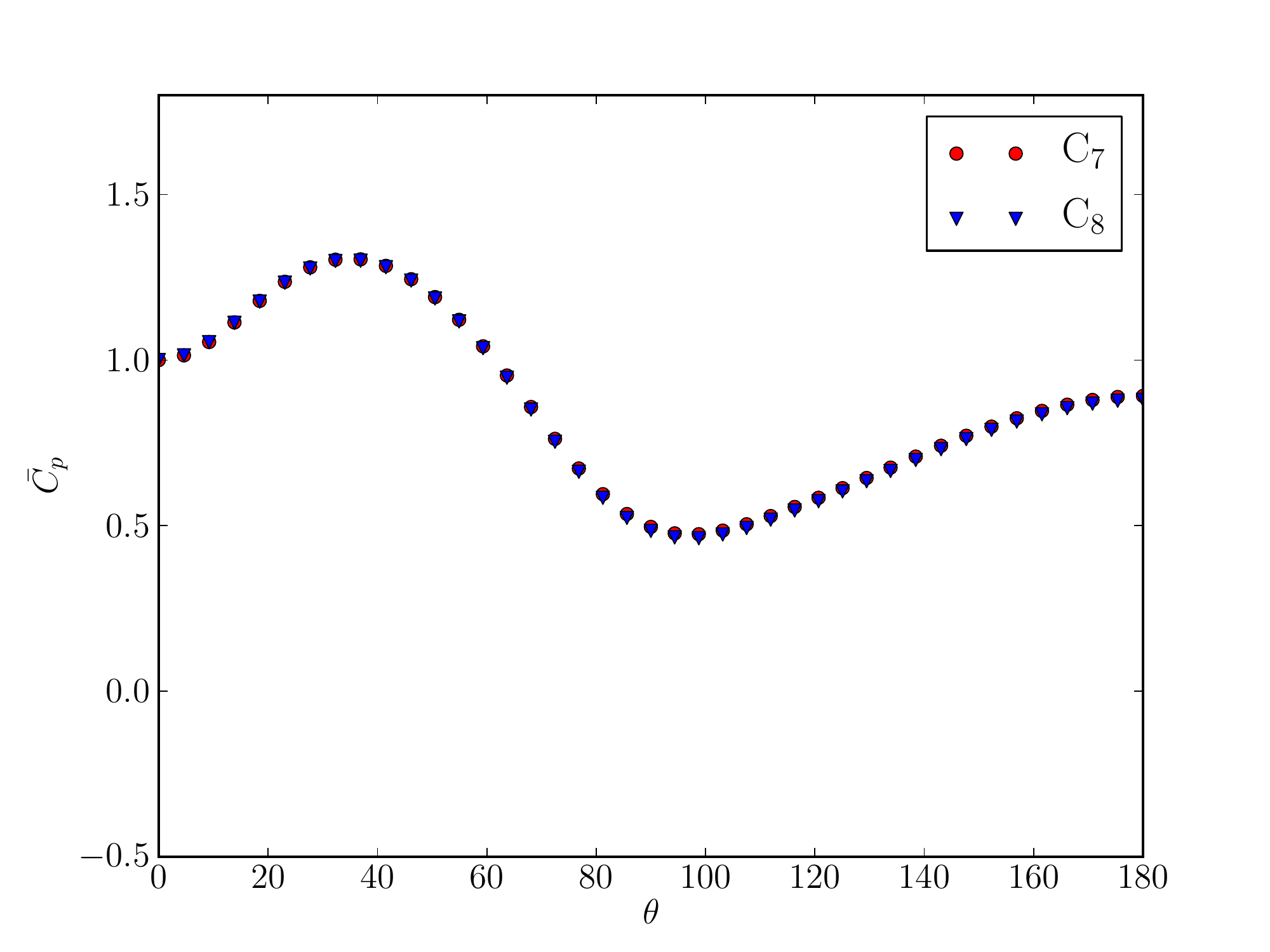}
\label{fig:subfig4_Cp_be}}
\caption[Mean pressure distribution on the middle cylinder
  surface. Definition of $\overline{C}_p$
  based on \cite{shim1988fluctuating}. (a) the third pair
  tube banks, (b) the fourth pair
  tube banks.]
{Mean pressure distribution on the middle cylinder
  surface. Definition of $\overline{C}_p$
  based on \cite{shim1988fluctuating}. (a) the third pair
  tube banks, (b) the fourth pair
  tube banks.}
\label{fig:globfig2_Cp_be}
\end{figure}

\begin{figure}[htp]
  \centering
  \includegraphics[trim=0.0in 0.0in 0.0in
  0.0in,clip=true,scale=1.0, angle = 0, width = 0.95\textwidth]{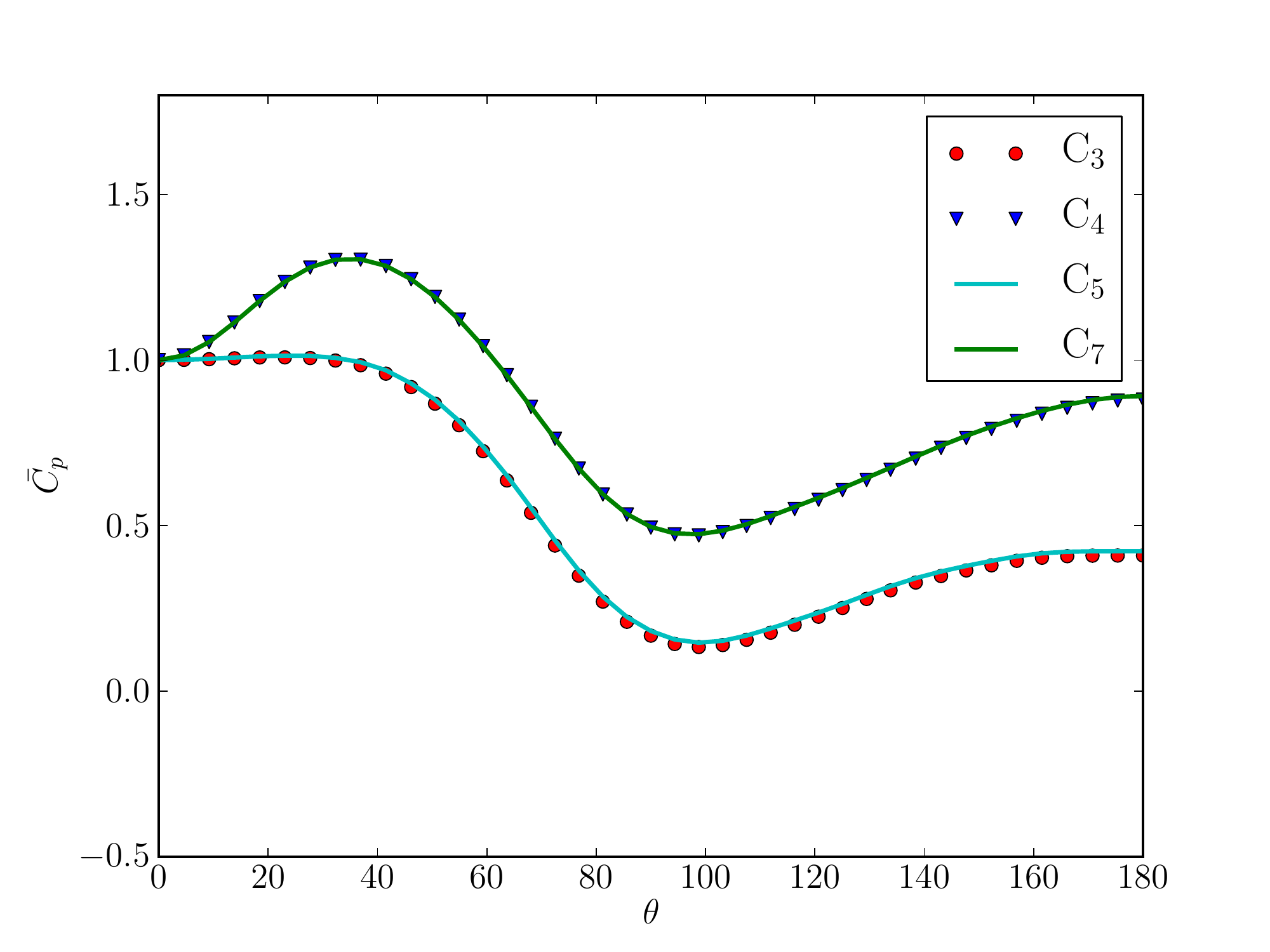} 
  \caption[Comparison of mean pressure distribution on the middle cylinder
  surface from the second, third and fourth pair of
  tube-banks]
{Comparison of mean pressure distribution on the middle cylinder
  surface from the second, third and fourth pair of
  tube-banks}
   \label{fig:be_tb_234}
\end{figure}

Figure~\ref{fig:globfig2_Cp_be} shows $\overline{C}_p$ on
the middle cylinders from the third and fourth pair of
tube-banks. Interestingly, there is no discernible
discrepancy between the two cylinders in the same pair.  For
the cylinder $C_5, C_6$, $\overline{C}_p$ behaviors like a
single cylinder.  Similar results on $\overline{C}_p$ are
observed for the cylinder $C_7, C_8$, but they are higher
than unit in the front side of cylinders. This is further
confirmed by the figure~\ref{fig:be_tb_234} which shows no
discrepancy between the $\overline{C}_p$ on cylinders from
different pairs.  This implies that the shedding vortex from
the third pair of tube-banks impinges on the final pair of
tube banks. This impingement may result in different surface
pressure distributions on cylinder $C_7, C_8$, but after the
scaling based on the equation:~\ref{eq:cp_ch7}
$\overline{C}_p$ displays no significant difference.

\subsection{Results on the particle phase}
\subsubsection{Sample particle trajectories and bounce upon impact}
\begin{figure}[htp]
  \centering
  \includegraphics[trim=0.0in 0.0in 0.0in
  0.0in,clip=true,scale=1.0, angle = 0, width = 0.95\textwidth]{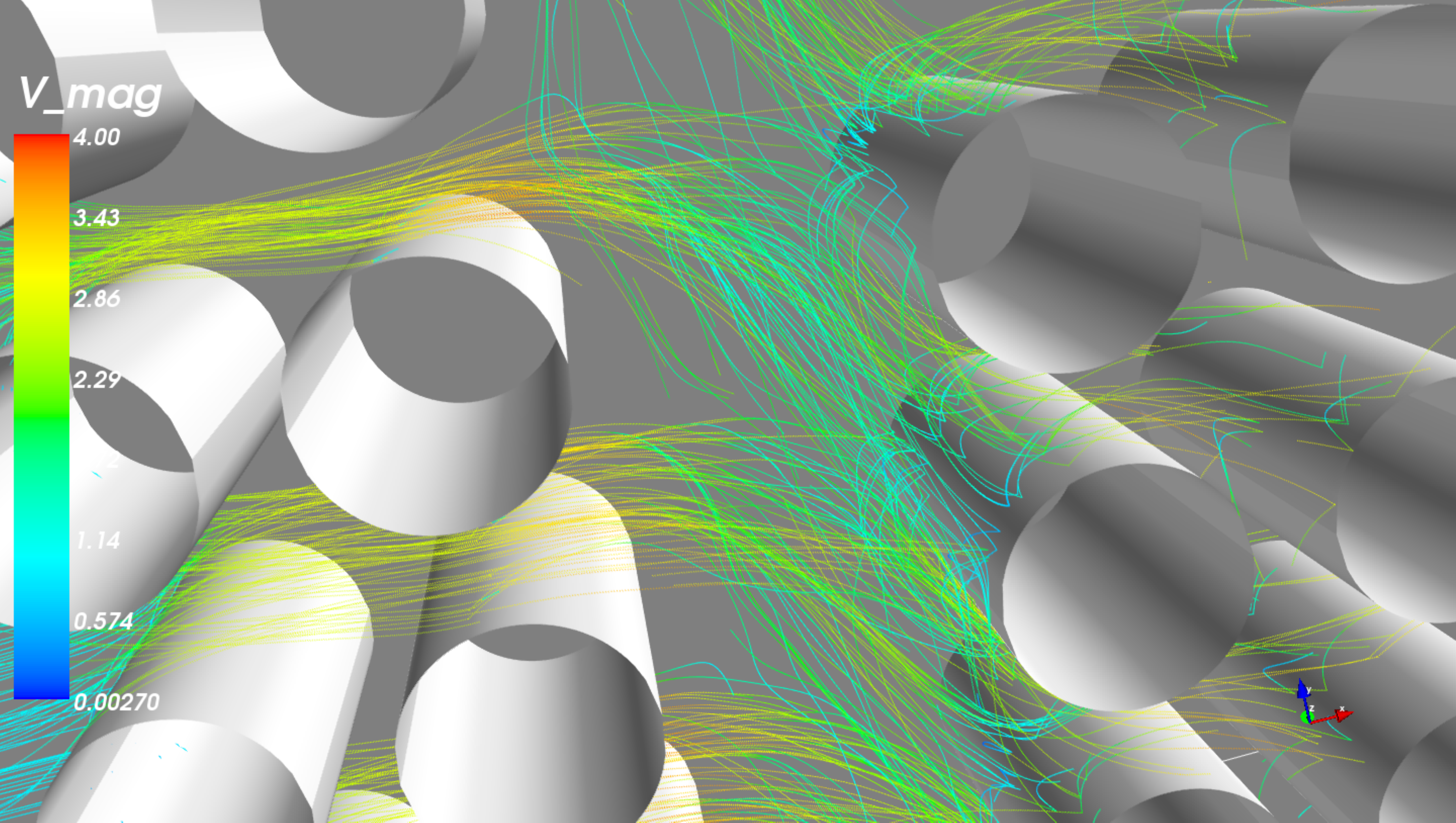} 
  \caption[Sample particle trajectories and bounce upon
  impact on cylinders for particles $St = 0.35$]
{Sample particle trajectories and bounce upon
  impact on cylinders for particles $St = 0.35$}
   \label{fig:sample_trj_tb}
\end{figure}

Figure~\ref{fig:sample_trj_tb} shows some sample particle
trajectories across the tube-banks. With the present
particle-wall collision model, it can be clearly observed
that some particles rebound upon
impact on the cylinders. This normally results in a smaller
rebounce velocity even particles peel on the surface of
cylinders as a result of energy loss.

\subsubsection{Deposition efficiency on tube-banks}
The deposition efficiency for a single cylinder(or known
as collection efficient) is normally defined as
\beq
\eta_{sc} = \frac{N_{dep}}{N_{tot}},
\eeq
where $N_{dep}$ is the number of deposition particles
on the cylinder, and $N_{tot}$ is the number of uniformly
distributed  particles in the upstream cross-sectional area
of the cylinders.

\begin{figure}[htp]
  \centering
  \includegraphics[trim=0.0in 0.0in 0.0in
  0.0in,clip=true,scale=1.0, angle = 0, width = 0.95\textwidth]{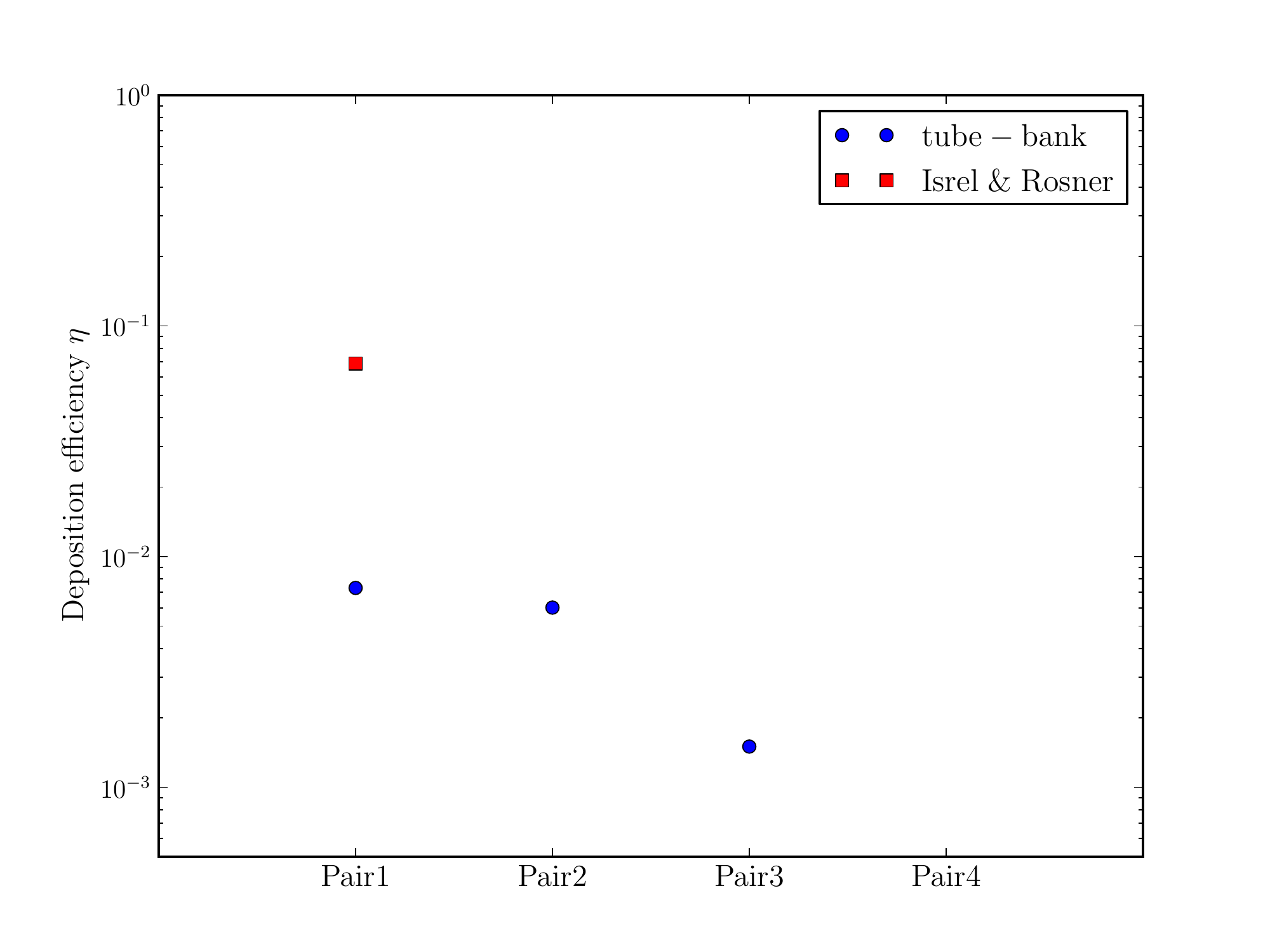} 
  \caption[Deposition efficiency of particles $St = 0.35$
  across tube-banks and a comparison against to the results
  on a single cylinder from \cite{israel1982use}]
{Deposition efficiency of particles $St = 0.35$
  across tube-banks and a comparison against to the results
  on a single cylinder from \cite{israel1982use}}
   \label{fig:dep_be_p2}
\end{figure}

\begin{figure}[htp]
  \centering
  \includegraphics[trim=0.0in 0.0in 0.0in
  0.0in,clip=true,scale=1.0, angle = 0, width =
  0.95\textwidth]{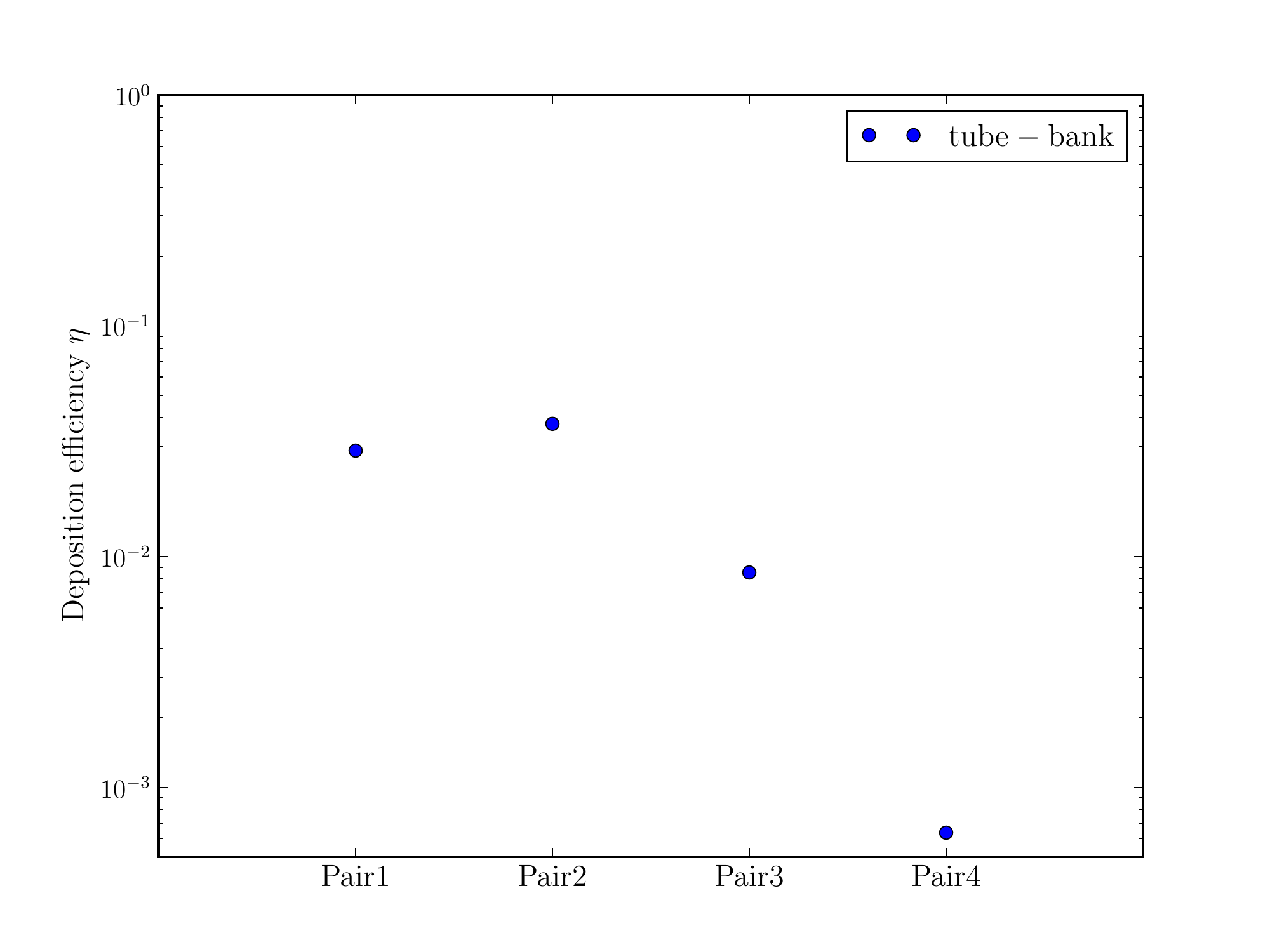}
\caption[Deposition efficiency of particles $St = 0.086$
  across tube-banks]
{Deposition efficiency of particles $St = 0.086$
  across tube-banks}
   \label{fig:dep_be_p1}
\end{figure}

\begin{figure}[htp]
  \centering
  \includegraphics[trim=0.0in 0.0in 0.0in
  0.0in,clip=true,scale=1.0, angle = 0, width =
  0.95\textwidth]{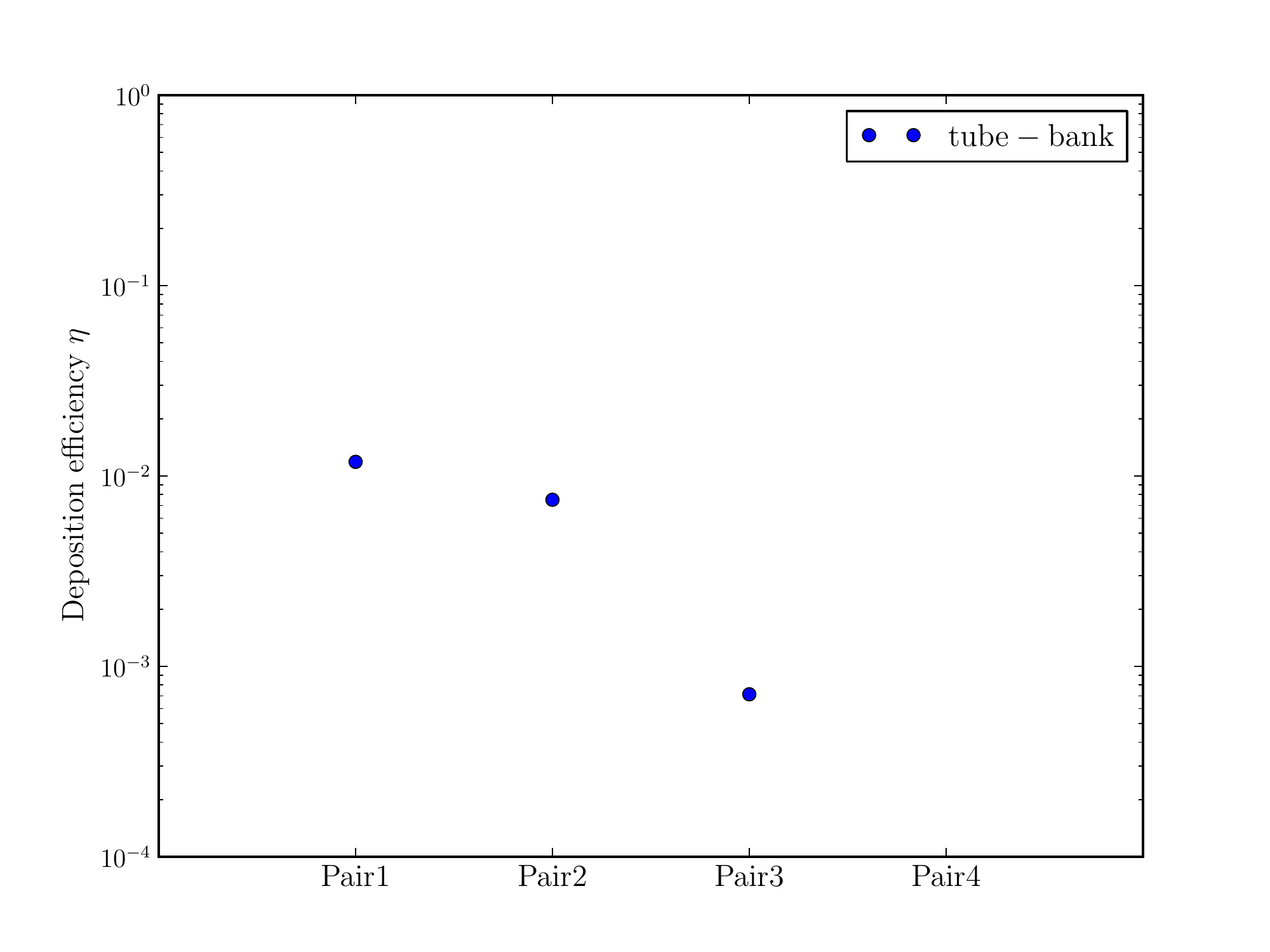}
 \caption[Deposition efficiency of particles $St = 0.0075$
  across tube-banks]
{Deposition efficiency of particles $St = 0.0075$
  across tube-banks}
   \label{fig:dep_be_p0}
\end{figure}

Nevertheless, in the present case that particles deposit on
in-line tube-banks, overall deposition efficiencies for each
pair of tube-banks have to be defined differently. The
deposition efficiency for the first pair is determined by
taking the number of particles in the upstream
cross-sectional area of the first column cylinders and comparing that
number to the number of particles actually deposited on the
first pair of tube-banks. This reads
\beq
\eta_{pair1} = \frac{N_{pair1}}{\frac{5}{6.94}N_{tot}}, 
\eeq 
where $N_{pair1}$ is the number of deposition particles on
the first pair of tube-banks, $N_{tot}$ is the total number
released from the upstream, $5/6.94$ is ratio of the
cross-sectional area of the first column cylinders to the
cross-sectional area of the computational domain. However,
since it is difficult to define how many particles are in
the upstream cross-sectional area of the succeeding
tube-banks, the number is assumed to be simply the number in
the particle release plane cross-sectional area
$5N_{tot}/6.94$ minus the number of particles deposited on
the preceding tube-banks. For example, the deposition
efficiency of the fourth pair of tube-banks can be written as
\beq
\eta_{pair_{4}} =
\frac{N_{pair_{4}}}{\Big(\frac{5}{6.94}N_{tot} -
    \sum_{i=1}^{3} N_{{pair_{i}}}\Big)}.
\eeq
The computed results on deposition efficiency across the
tube-banks are shown in figure~\ref{fig:dep_be_p2},
\ref{fig:dep_be_p1} and \ref{fig:dep_be_p0}.

Figure~\ref{fig:dep_be_p2} also includes a result on
deposition efficiency on a single cylinder based on the
curve fit proposed by \citeauthor{israel1982use}\cite{israel1982use}. The particle $St
= 0.35$ an effective Stokes number $St_{eff}$ equalling to
$0.21$ based on equation ~\ref{eq:st_psi}, which is in the
valid range of above curve-fit. First, it can be seen that
the deposition efficiency for $St = 0.35$ is falling along
the flow direction. Second, the deposition efficiency for
the first pair of tube-banks is significantly lower than
that of a single circular cylinder. This may result from the
fact that flow channeling is occurring between two vertically
adjacent cylinders. However, it can be observed that there
is a considerable increase of the deposition efficiency for
particles $St = 0.086, 0.0075$, which is not consistent with
theoretical results on a single cylinder. A possible
explanation is that a large amount of particles are
entrained into the wake of the back-banks of each pair of
tube-banks and get deposition by back-side
impaction~(see~\cite{haugen2010particle}).

\subsubsection{Deposition fraction across the tube-banks}
Deposition data of three sets of particles ($St = 0.35,
0.086, 0.0075$) on tube-banks are presented here. The pair
of tube-banks shown in figure~\ref{fig:be_tubebank} are
designated by $\mathrm{pair1}$ (upstream) through
$\mathrm{pair4}$ (downstream); the upstream and downstream
column of tube-banks in each pair are designated by front
bank and back bank.  $10^7$ particles are released into the
computational continuously for an interval of $1000$
continuous time steps, in order to both collect enough
particles and account for the unsteady effect.

\begin{figure}[htp]
\centering
\subfloat[][]{
\includegraphics[width=0.95\textwidth]{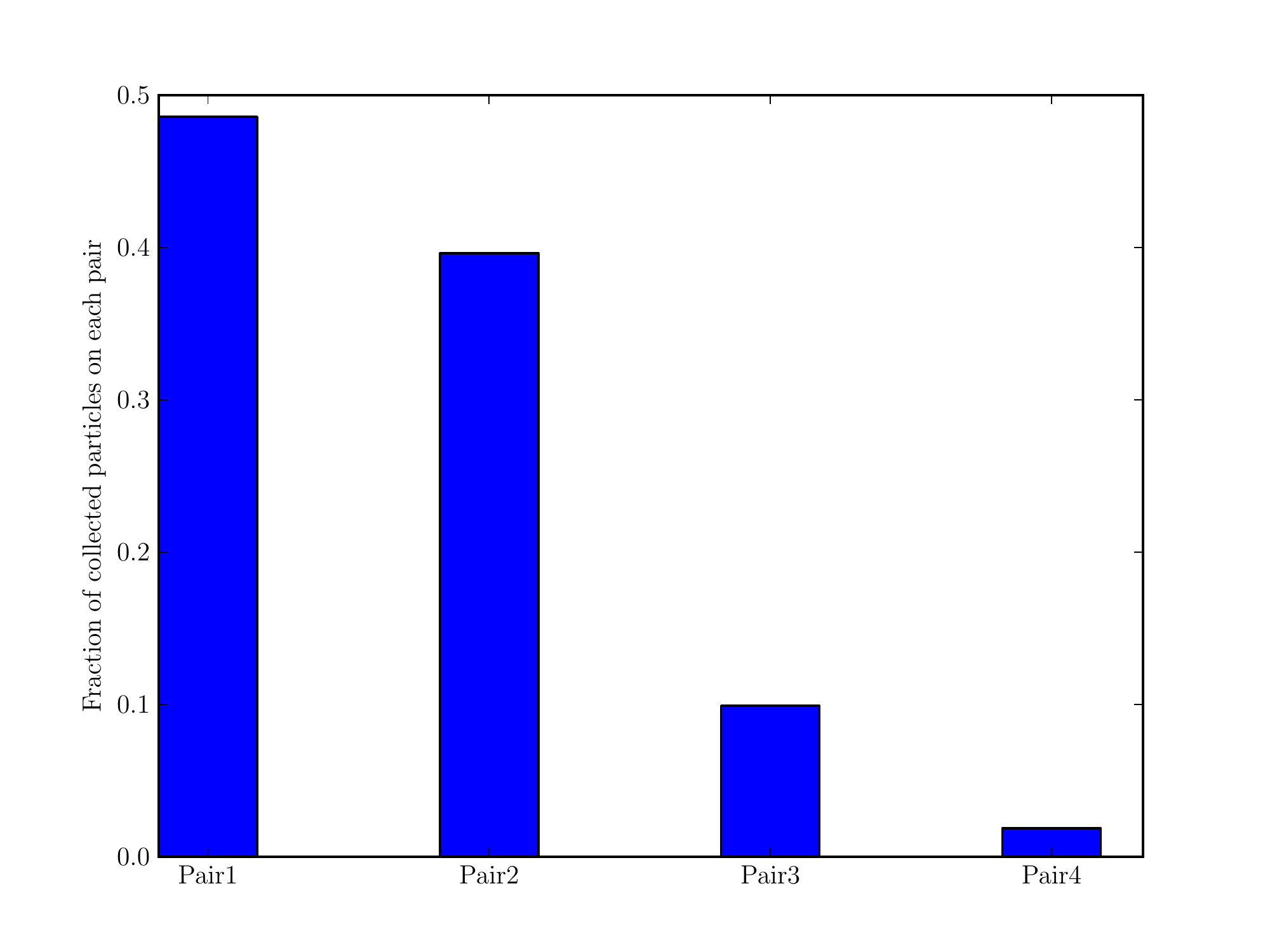}
\label{fig:subfig1_be_dep}}\\
\subfloat[][]{
\includegraphics[width=0.95\textwidth]{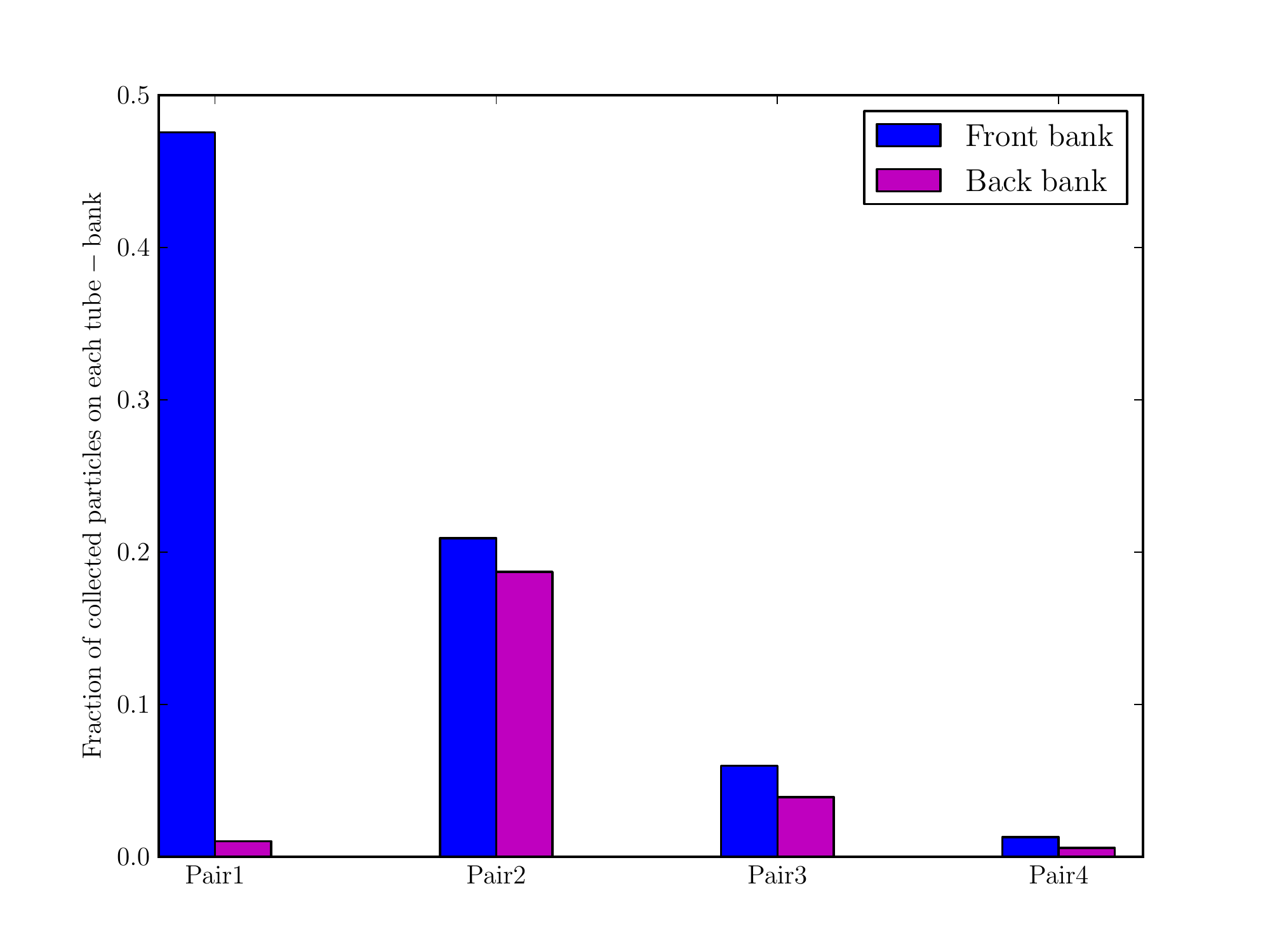}
\label{fig:subfig2_be_dep}}
\caption[Fraction of total deposition particles across
tube-banks for particles $St = 0.35$ (a) each
  pair of tube-banks, (b) each tube-banks.]
{Fraction of total deposition particles across tube-banks
  for particles $St = 0.35$ (a) each
  pair of tube-banks, (b) each tube-banks.}
\label{fig:globfig1_be_dep}
\end{figure}

\begin{figure}[htp]
\centering
\subfloat[][]{
\includegraphics[width=0.95\textwidth]{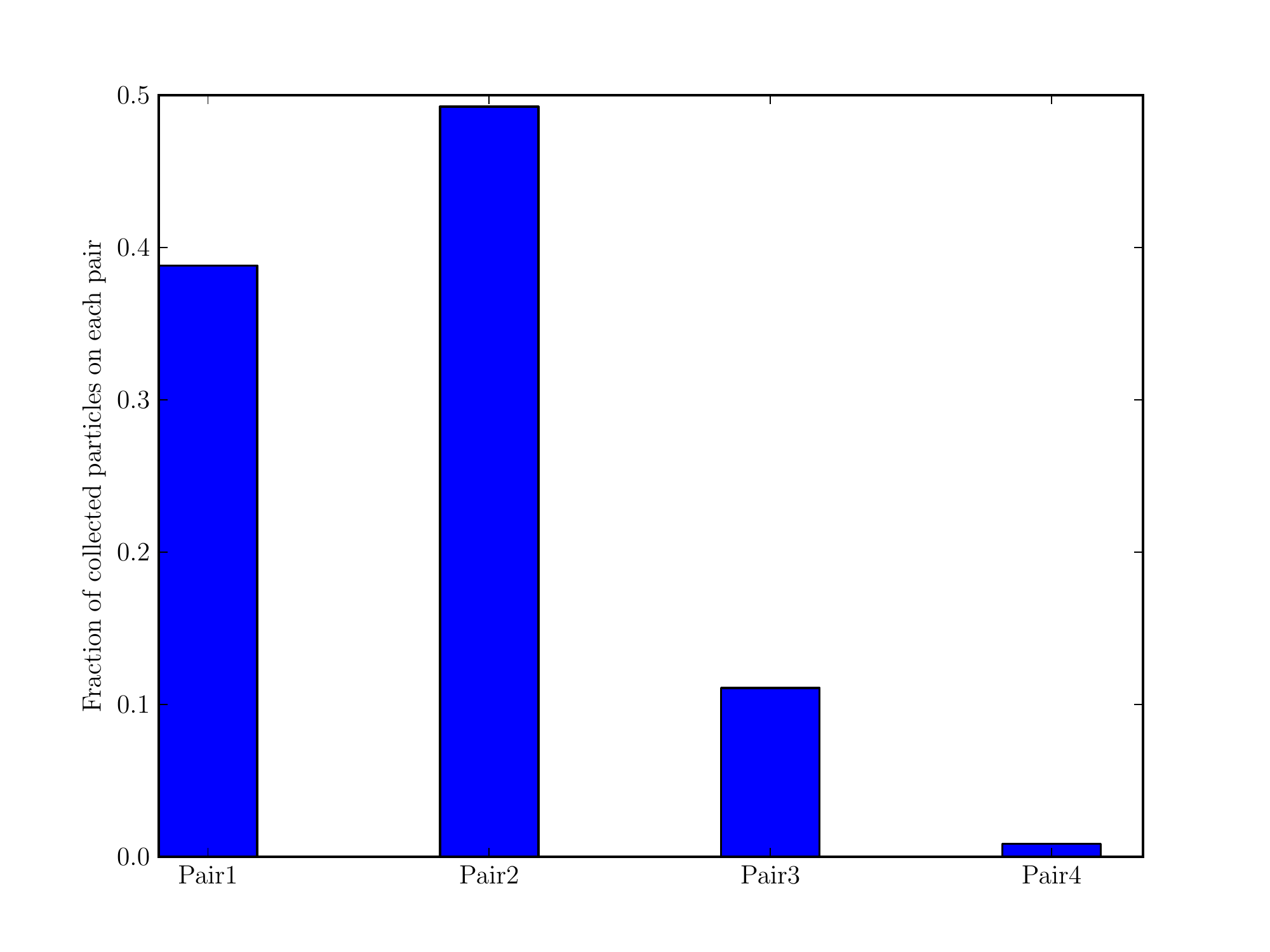}
\label{fig:subfig3_be_dep}}\\
\subfloat[][]{
\includegraphics[width=0.95\textwidth]{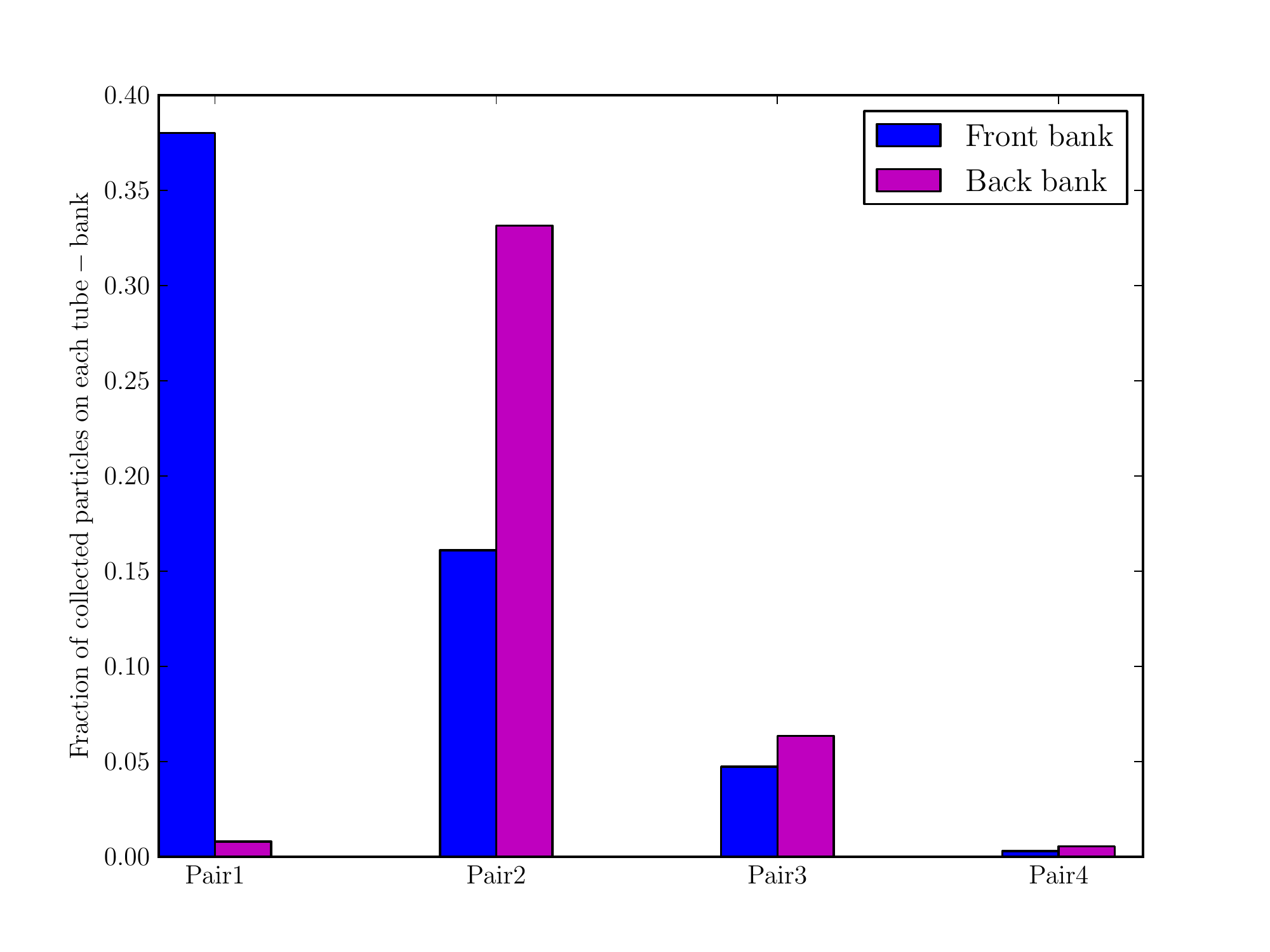}
\label{fig:subfig4_be_dep}}
\caption[Fraction of total deposition particles across
tube-banks for particles $St = 0.086$ (a) each
  pair of tube-banks, (b) each tube-banks.]
{Fraction of total deposition particles across tube-banks
  for particles $St = 0.086$ (a) each
  pair of tube-banks, (b) each tube-banks.}
\label{fig:globfig2_be_dep}
\end{figure}

\begin{figure}[htp]
\centering
\subfloat[][]{
\includegraphics[width=0.95\textwidth]{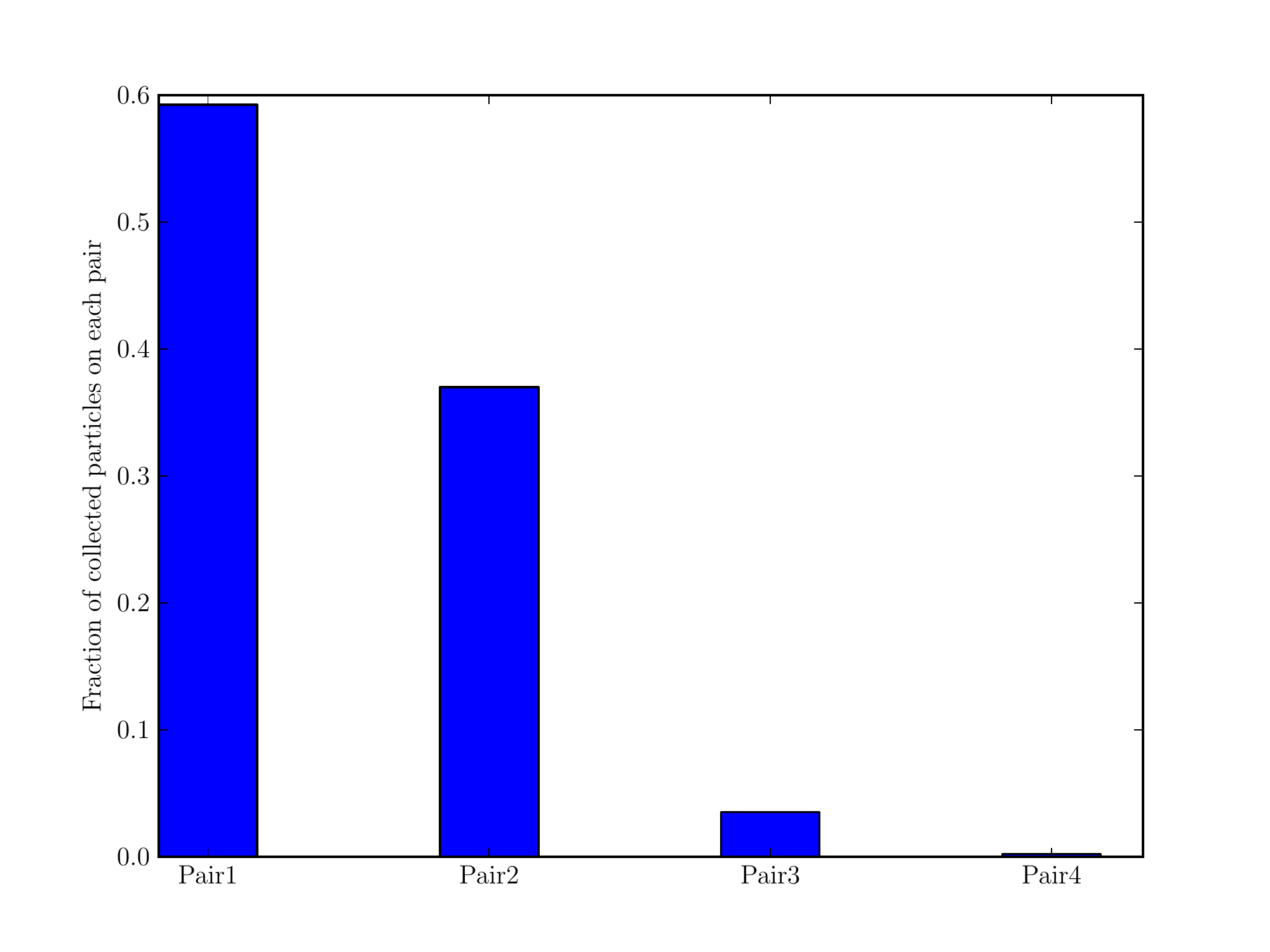}
\label{fig:subfig5_be_dep}}\\
\subfloat[][]{
\includegraphics[width=0.95\textwidth]{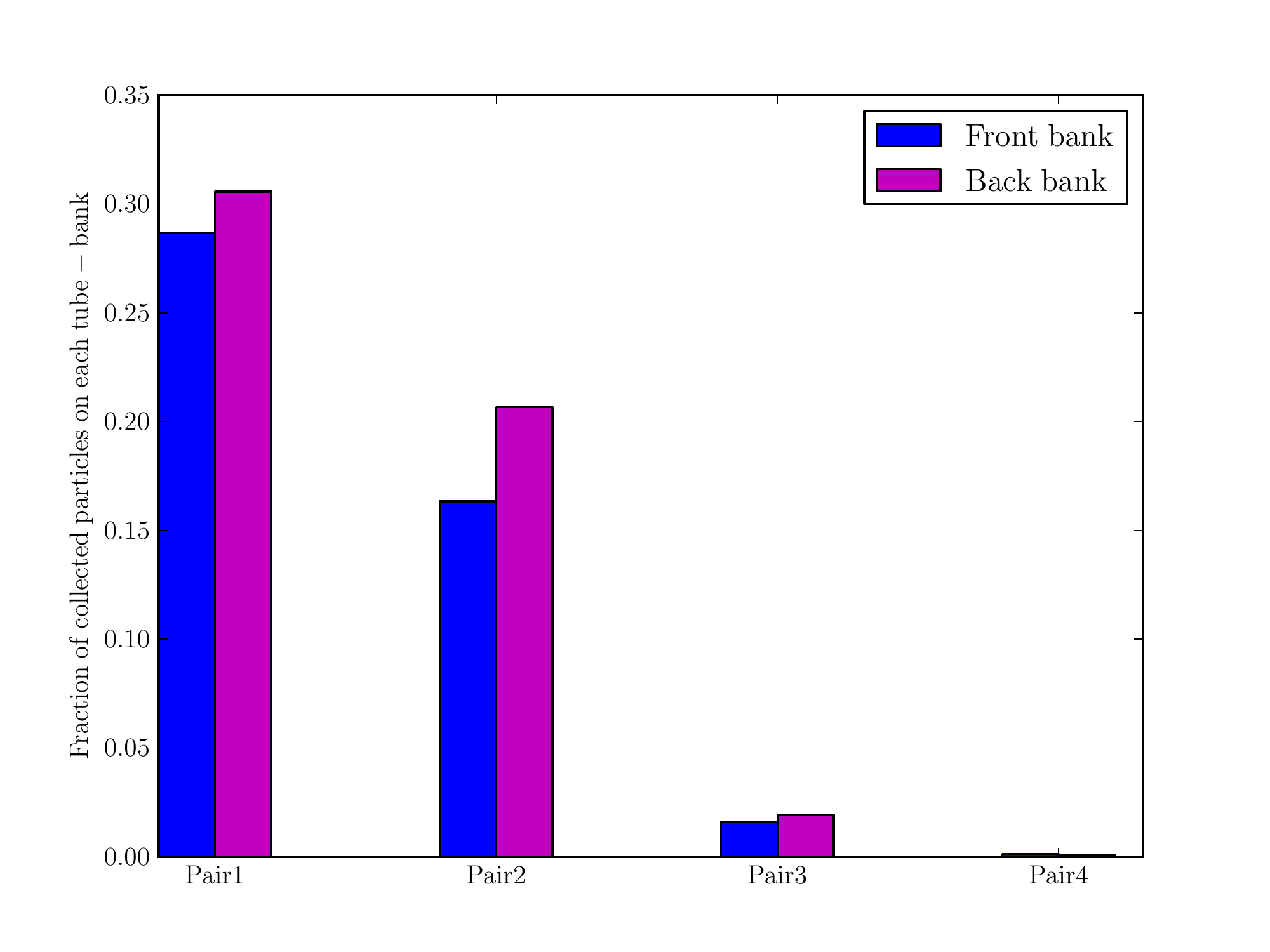}
\label{fig:subfig6_be_dep}}
\caption[Fraction of total deposition particles across
tube-banks for particles $St = 0.0075$ (a) each
  pair of tube-banks, (b) each tube-banks.]
{Fraction of total deposition particles across tube-banks
  for particles $St = 0.0075$ (a) each
  pair of tube-banks, (b) each tube-banks.}
\label{fig:globfig3_be_dep}
\end{figure}

Figure~\ref{fig:globfig1_be_dep}, \ref{fig:globfig2_be_dep}
and \ref{fig:globfig3_be_dep} show that the variations of
fraction of total deposition particles across the whole
tube-banks for the three sets particles. As can be seen, for
the particle $St = 0.35$, across the four pairs of the
tube-banks the fraction of deposition particles on the pair1
is significantly higher than on the downstream pair of
tube-banks, pair2, pair3 and pair4. On the other hand, the
fraction of deposition on the front bank in the same pair of
tube-banks is higher than on the back bank, especially for
the pair1.  This may result from the fact that a significant
part of this set of particle is entrained in the bulk flow
between cylinders and don't follow the shedding vortex from
the preceding cylinders (see
figure~\ref{fig:sample_trj_tb}). However, the computed
results for particles $St = 0.086, 0.0075$ are not
consistent with those results for $St = 0.345$. Although the
fraction of deposition on the pair3 and pair4 are lower than
the preceding two pairs, a striking difference is noted when
compared to the particle $St = 0.35$.

\begin{figure}[htp]
\centering
\subfloat[][]{
\includegraphics[trim=1.8in 0.0in 1.0in
  0.0in,clip=true,scale=1.0, angle = 0, width = 0.95\textwidth]{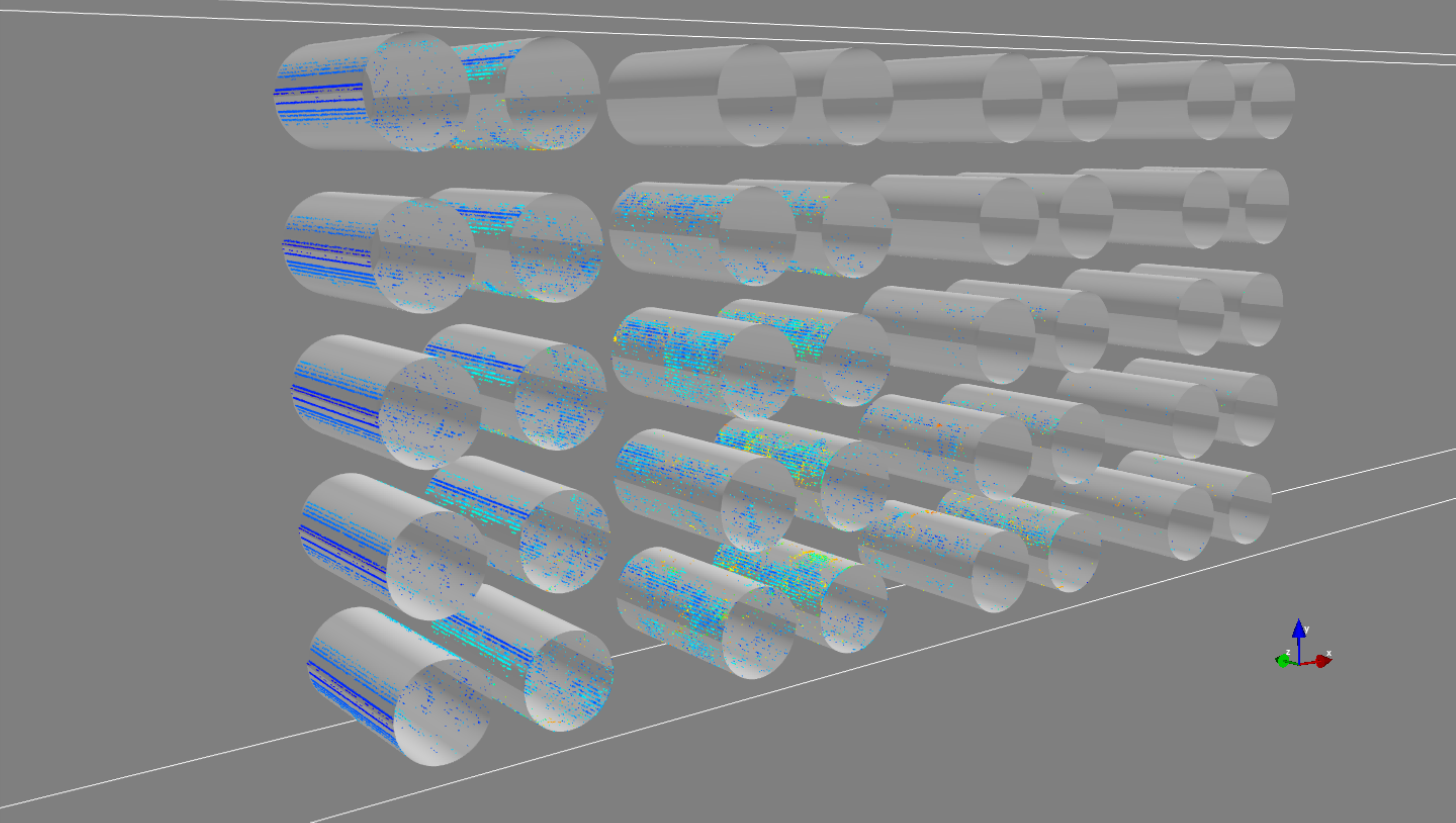}
\label{fig:subfig1_be_dep_p}}\\
\subfloat[][]{
\includegraphics[trim=1.0in 0.0in 1.0in
  0.0in,clip=true,scale=1.0, angle = 0, width =
  0.95\textwidth]{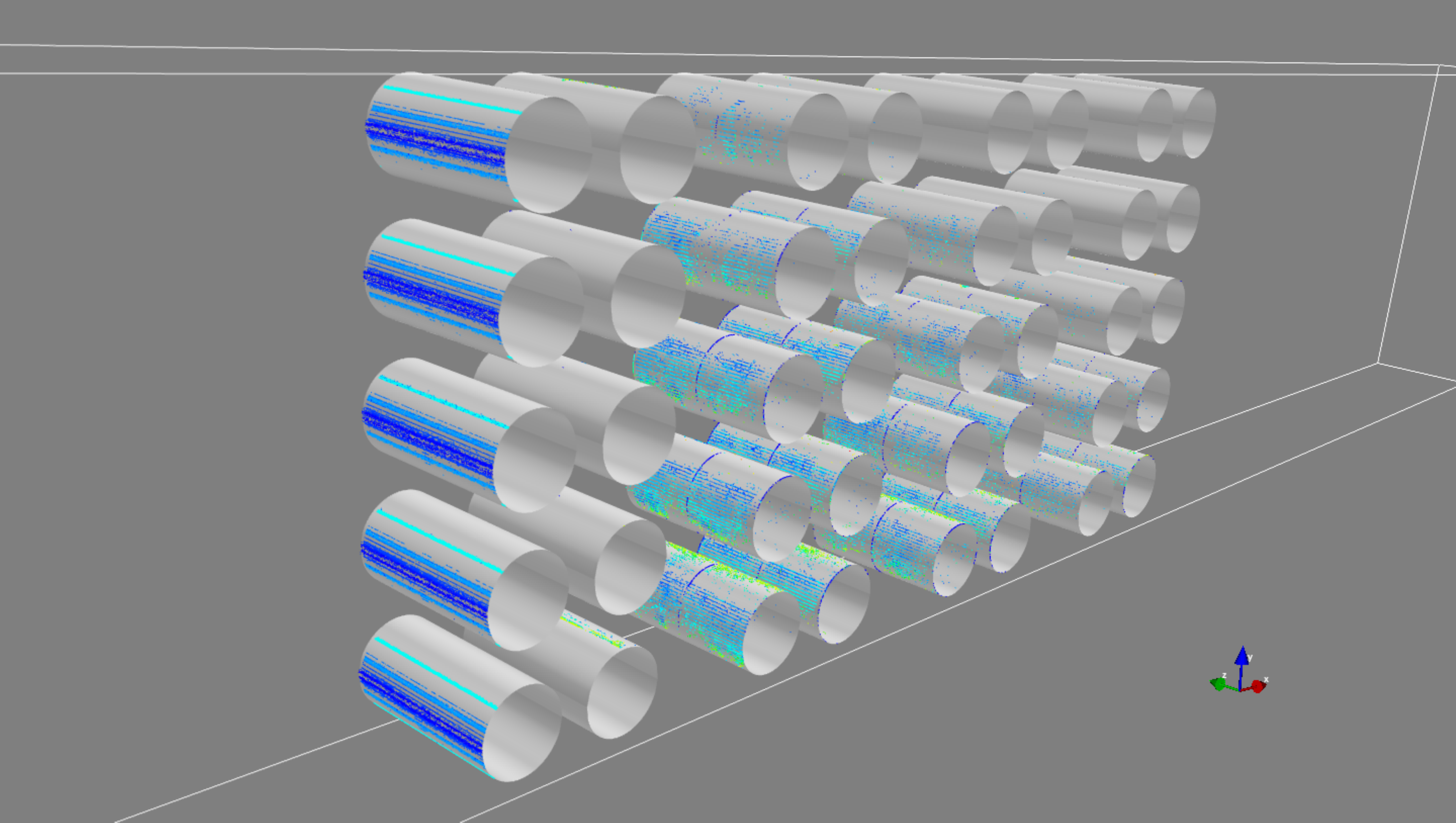}
\label{fig:subfig2_be_dep_p}}
\caption[Deposition particles on
tube-banks (a)  $St = 0.0075$, (b)  $St = 0.35$.]
{Deposition particles on
tube-banks (a)  $St = 0.0075$, (b)  $St = 0.35$.}
\label{fig:globfig1_be_dep_p}
\end{figure}

It can be observed from figure~\ref{fig:subfig1_be_dep_p}
that the back-side deposition on the
back banks of pair1 and pair2, for particles $St = 0.0075$
result in considerably higher fraction of deposition. For
the back-side deposition, \citeauthor{haugen2010particle}\cite{haugen2010particle} argued
that when particle with response time $\tau_p$ is close to
the eddy time $\tau_{eddy}$, they normally follow the
eddies in the wake of the tube-banks and gain enough
momentum to impact on the cylinders.

\section{Concluding remarks}
A large eddy simulation of inertia deposition, impact of
heavy particles on an in-line tube-banks in a turbulent flow
are performed. The flow Reynolds number, based on the
cylinder diameter $D$ and flow velocity between the gap of
two vertically adjacent cylinders, is $33960$. Flow
structures across the tube bank based on the normalized Q
criterion is presented. Following the formula for mean
pressure distribution for cylinders in tube-banks proposed
by \citeauthor{shim1988fluctuating}\cite{shim1988fluctuating}, mean pressure distribution on
the middle cylinder from each tube-bank is present. The $S$
shape of mean pressure distribution is observed on the
back-bank of the first pair and second pair of
tube-banks. Further, the mean pressure distribution on the
each tube-bank within the third pair of tube-banks displays
almost exactly the same behavior. Similar behavior is
observed on the final pair of tube-banks.

The results for three sets of particles ($St = 0.35, 0.086,
0.0075$) are based on $10^7$ particles tracked through
Lagrangian point particle tracking approach. The particle
bounce upon impact is taken into account through a
particle-wall collision model. Some sample particle
trajectories across tube-banks are shown, which shows that
some particles rebound from the cylinder surface upon
impact.  The deposition efficiency for the three sets of
particles are presented across the tube-banks. Moreover, the
fraction deposition of particles across each tube-banks are
provided. It is observed that for the particle $St = 0.35$
most of particles get deposition on the first tube-banks,
especially on the first column. This is consistent with
practical experience that the first column of tube-banks
plays a protection role on mitigating fouling on the
succeeding tube-banks. Based on the effective Stokes number
proposed by \citeauthor{israel1982use}\cite{israel1982use}, the overall deposition
efficiency of the particle $St = 0.35$ on the first pair of
tube-banks is significantly lower than that of a single
circular cylinder. Nevertheless, the results on deposition
efficiency for particles $St = 0.086, 0.0075$ are not
consistent with those of particles $St = 0.35$. A display of
deposition particle on tube-banks suggests that much more of
smaller particles get deposition on the back-side of the
back-banks. This may due to the fact that the smaller
particles are easier to be entrained into the wake and
impact onto the back-side of cylinders. This issue needs
further investigations.

\section{acknowledgments}
We wish to acknowledge the support of British Energy (Part
of EDF).

\clearpage

\bibliographystyle{model1-num-names}
\bibliography{cJin_jp4_jas}

\end{document}